\documentclass[twocolumn,amssymb, nobibnotes, showpacs, aps, pra,10pt]{revtex4-1}
\pdfoutput=1
\usepackage{amsmath,amssymb,graphicx}
\usepackage{eucal}
\usepackage{braket}
\usepackage{mathtools}
\usepackage{siunitx}
\usepackage{xcolor}
\usepackage{subfigure}
\usepackage{hyperref}

\DeclarePairedDelimiterX{\norm}[1]{\lVert}{\rVert}{#1}
\newcommand*{\ai}{\hat{a}_{\textrm{i}}}
\newcommand*{\as}{\hat{a}_{\textrm{s}}}
\newcommand*{\aione}{\hat{a}_{\textrm{i}1}}
\newcommand*{\aitwo}{\hat{a}_{\textrm{i}2}}
\newcommand*{\abone}{\hat{a}_{\textrm{b}1}}
\newcommand*{\abtwo}{\hat{a}_{\textrm{b}2}}

\newcommand*{\asone}{\hat{a}_{\textrm{s}1}}
\newcommand*{\astwo}{\hat{a}_{\textrm{s}2}}
\newcommand*{\wione}{\omega_{\textrm{i}1}}
\newcommand*{\wi}{\omega_{i}}
\newcommand*{\ws}{\omega_{s}}
\newcommand*{\wpbar}{\bar{\omega}_{\text{p}}}
\newcommand*{\witwo}{\omega_{\textrm{i}2}}
\newcommand*{\waone}{\omega_{\textrm{a}1}}
\newcommand*{\watwo}{\omega_{\textrm{a}2}}
\newcommand*{\wbone}{\omega_{\textrm{b}1}}
\newcommand*{\wbtwo}{\omega_{\textrm{b}2}}

\newcommand*{\wsone}{\omega_{\textrm{s}1}}
\newcommand*{\wstwo}{\omega_{\textrm{s}2}}
\newcommand*{\wsfg}{\omega_\textrm{SFG}}

\newcommand*{\asfg}{\hat{a}_\textrm{SFG}}
\newcommand*{\vacuum}{\ket{\textrm{vac}}}
\newcommand*{\Pavg}{P_{\text{avg}}}
\newcommand*{\hc}{\textrm{H.c.}}

\newcommand*{\ii}{\mathrm{i}}

\newcommand*{\ie}{i.e.\ }
\newcommand*\diff{\mathop{}\!\mathrm{d}}
\DeclareMathOperator*{\trace}{\textrm{Tr}}
\DeclareMathOperator*{\sinc}{\textrm{sinc}}

\begin{document}

\title{Entanglement Swapping for Generation of Heralded Time-Frequency-Entangled Photon Pairs}

\author{Dashiell L. P. Vitullo}
\affiliation{Department of Physics and Oregon Center for Optical, Molecular, \& Quantum Science, University of Oregon, Eugene, Oregon 97403, USA}

\author{M. G. Raymer}
\affiliation{Department of Physics and Oregon Center for Optical, Molecular, \& Quantum Science, University of Oregon, Eugene, Oregon 97403, USA}

\author{B. J. Smith}\email{Corresponding author: bjsmith@uoregon.edu}
\affiliation{Department of Physics and Oregon Center for Optical, Molecular, \& Quantum Science, University of Oregon, Eugene, Oregon 97403, USA}

\author{Micha\l{} Karpi\'nski}
\affiliation{Faculty of Physics, University of Warsaw, Pasteura 5, 02-093 Warszawa, Poland}

\author{L. Mejling}
\affiliation{Department of Photonics Engineering, Technical University of Denmark, 2800 Kgs.~Lyngby, Denmark}
\author{K. Rottwitt}
\affiliation{Department of Photonics Engineering, Technical University of Denmark, 2800 Kgs.~Lyngby, Denmark}

\begin{abstract}
Photonic time-frequency entanglement is a promising resource for quantum information processing technologies. We investigate swapping of continuous-variable entanglement in the time-frequency degree of freedom using three-wave mixing in the low-gain regime with the aim of producing heralded biphoton states with high purity and low multi-pair probability. Heralding is achieved by combining one photon from each of two biphoton sources via sum-frequency generation to create a herald photon. We present a realistic model with pulsed pumps, investigate the effects of resolving the frequency of the herald photon, and find that frequency-resolving measurement of the herald photon is necessary to produce high-purity biphotons. We also find a trade-off between the rate of successful entanglement swapping and both the purity and quantified entanglement resource (negativity) of the heralded biphoton state. 
\end{abstract}



\maketitle 

\section{Introduction}

Entangled photon pairs are an important resource for quantum communication, quantum metrology, and quantum networking \cite{Gisin2007,Torres2011}. Their generation is typically not deterministic, for either fundamental reasons, as in sources based on spontaneous nonlinear optical processes, or for technical reasons, as is the case of single quantum emitters, such as quantum dots or atoms, where losses arise from imperfect coupling of photon pairs into the desired optical modes \cite{Takeuchi2013}. The optical-field state generated by these sources necessarily contains an undesired vacuum component, and additional unwanted multi-pair components are present for spontaneous nonlinear optical sources. Sources that deterministically generate exactly the desired number of entangled photon pairs would enable large-scale quantum information processing and would be an important resource for secure quantum communication.

In the absence of true deterministic entangled pair generation, the heralding approach, where a pair generation event is heralded by an accompanying signal, enables efficient realization of quantum operations \cite{Barz2010}. Heralding removes the vacuum component from the optical-field state at the cost of reduced generation probability and can be implemented in ways that remove higher-order components that contain more than the desired number of photon pairs.

Entanglement swapping has been proposed as a means to convert two nondeterminalistically generated photon pairs into a single heralded entangled photon pair \cite{Yurke1992, Zukowski1993}. In this scheme, two independent nondeterministic sources each create an entangled photon pair. One photon of each pair is in a spatial mode which we will call {\em active}, and the other in the {\em bystander} mode.  The active modes from each source are jointly measured, and the measurement result indicates whether or not entanglement has been successfully swapped. A measurement result indicating successful swapping heralds creation of entanglement between the remaining pair of bystander photons, and indicates that the swapping process has erased information about the state of the converted active photons in the entangled degree of freedom. Bystander photons prepared in this manner are entangled despite having never been in the same place at the same time. Entanglement swapping is particularly relevant for spontaneous parametric downconversion (SPDC) and spontaneous four wave mixing (SFWM) sources, which are inherently probabilistic \cite{Torres2011}, and has been analyzed and demonstrated in many degrees of freedom of the optical field \cite{Jennewein2001, Riedmatten2005,Takei2005,Kaltenbaek2009,Takesue2009, Sangouard2011,Jin2015, Zhang2016}.

The spectral-temporal degree of freedom of light has been recently recognized as a promising framework for quantum information science, since it enables multidimensional encoding of quantum information in a way compatible with existing guided-wave and free-space optical infrastructure \cite{Humphreys2013,Nunn2013,Roslund2014,Kowligy2014,Donohue2014,Brecht2015,Lukens2017,Wright2017,Karpinski2017,Kues2017,Villegas2017,FiberEntanglement}.   
Moreover, spectral-temporal entanglement naturally arises in SPDC and SFWM as a consequence of energy conservation in parametric optical processes.

\begin{figure*}[t!]
\centerline{\includegraphics[width=0.9\textwidth]{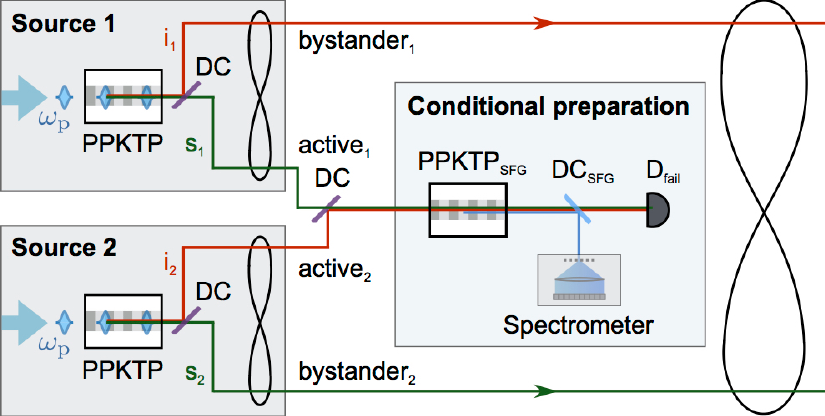}}
\caption{Entanglement swapping setup. The infinity symbols denote entangled photon pairs. Pulsed pump beams, denoted in blue, are directed into PPKTP-waveguide-based SPDC sources.  The active fields comprise the signal from source 1 ($s_1$) and idler from source 2 ($i_2$), while the remaining $i_1$ and $s_2$ are the bystander fields. Dichroic filters (DC) separate the signal and idler fields within the sources, and combine the active fields. $\text{PPKTP}_{\text{SFG}}$ is the SFG PPKTP waveguide and $\text{DC}_{\text{SFG}}$ is the dichroic filter separating the sum-frequency converted photons from unconverted active photons. Entanglement swapping is performed when exactly one SFG photon is detected with an ideal spectrometer and there is no detection at the fail detector $D_{\mathrm{fail}}$, (a single-photon counting module).} \label{fig:setup}
\end{figure*}

Realization of spectral-temporal entanglement swapping (STES) will provide another important tool for implementing quantum information processing using time-frequency modes (i.e.\ temporal modes). The necessary joint measurement of the active downconversion modes can be implemented using sum-frequency generation (SFG) in a nonlinear optical medium, as proposed by Molotkov and Nazin \cite{Molotkov1999}.  These authors analyzed STES using an idealized model in which the pump laser is monochromatic and the SFG phase-matching bandwidth is infinite.

In this paper, we analyze an experimentally realistic implementation of STES by SFG on photons generated by SPDC. Our design includes pulsed pump beams, which are necessary for clocked operation in quantum information processing networks, and realistic phase matching constraints, which crucially affect the joint measurement of the active modes.  We discuss the necessity of frequency-resolving herald detection and propose a design that produces high-purity entangled photon-pair states. We verify that the scheme not only creates heralded entangled pairs, but also that the heralded pair states contain greater entanglement than is present in the states produced by the SPDC sources. The scheme also suppresses multi-pair generation events. The major limitation of the method is the low heralding rate, although we point out that the needed joint measurement, performed using SFG of two individually generated single photons (without any additional pumping), has been demonstrated experimentally in \cite{Guerreiro2014}.

\section{Concept}
Our design for spectral-temporal entanglement swapping is presented in Fig.\ \ref{fig:setup}. Pulsed pump beams of central frequency $\omega_p$ are sent through periodically-poled potassium titanyl phosphate (PPKTP)  waveguides in crystals of length $L$ and poling period $\Lambda$ to create photon pairs via SPDC in the type-II phase matching configuration. Dichroic filters (DC) within each source separate the frequency-nondegenerate signal and idler fields. The active modes, composed of the signal from source 1 ($s_1$) and the idler from source 2 ($i_2$), are directed into a third PPKTP waveguide of length $L_{\text{SFG}}$ where the sum-frequency generation (SFG) process probabilistically combines them into a photon at the original pump frequency. This erases information about the difference frequency between the converted photons. A subsequent dichroic filter ($\text{DC}_\text{SFG}$) directs successfully converted light to an ideal spectrometer, to measure $\wsfg$. This heralds the generation of a spectrally entangled two-photon state between the bystander fields, $i_1$ and $s_2$, which have never interacted. The other output of $\text{DC}_\text{SFG}$ is monitored by a single-photon-counting module, constituting the ``fail'' detector, $D_{\text{fail}}$.   Simultaneous detection in the spectrometer and at $D_{\text{fail}}$ indicates that more than one pair-generation event took place in at least one source, and the bystander modes contain more photons than desired. Similarly, detection of more than one photon in the  spectrometer indicates the bystander modes have more than the desired number of photons. Thus, conditioning the use of the output photons on a successful SFG frequency measurement in a single bin and no detection at $D_{\text{fail}}$ prepares a heralded single photon-pair output state with entanglement between the bystander modes and substantially suppressed contributions from higher-order photon number terms.

It is important to consider whether the swapping process can distinguish two photon-pair creation events that occur in the same source from events that occur in separate sources. The creation process in source 1 is statistically independent of the process in source 2. Thus, during a given pump pulse, the probability that two photon-pair generation events will happen in a single source is the same as the probability that a single photon-pair generation event occurs in each of the two sources. If the active field photons are indistinguishable, then two pair-generation events in the same crystal would give rise to false herald detections (so-called because the output photons occupy the same field and are not usefully entangled) with probabilities comparable to those for true herald detections. In our design, the phase matching in the SFG crystal is satisfied only when both a signal and an idler photon (one each from two separate sources) are present. This avoids the false herald pitfall because two pair-generation events in the same crystal generate two photons that do not satisfy the phase mismatch requirement for generation of an SFG photon.
 
\section{Theory} \label{sec:theory}
A single spontaneous parametric down conversion (SPDC) source generates the state
\begin{equation}
\begin{split}
&\ket{\Psi} = \sqrt{1-|\xi|^2 - \mathcal{O}(|\xi|^4)}\vacuum \\
 + \xi \int_0^\infty &\text{d}\wi \text{d}\ws \bar{\Phi}(\wi, \ws) \ai^\dagger(\wi) \as^\dagger(\ws)  \vacuum + \mathcal{O}(\xi^2), \label{eq:SPDCstate}
\end{split}
\end{equation}
where $\xi$ is the probability amplitude for creating a photon-pair, the $s$ and $i$ subscripts denote the signal and idler fields respectively, and $\Phi(\wi, \ws) = \xi \bar{\Phi}(\wi, \ws)$ is the two-frequency joint spectral amplitude (JSA). We consider the low-gain regime where the probability of a single biphoton creation event $|\xi|^2 \ll 1$. The higher-order terms represented with order $\mathcal{O}(\xi^2)$ make non-negligible contributions to the state input to the swapping process, but we configure this process so these terms make negligible contributions to the state prepared upon herald detection. For clarity, we neglect terms of order $\mathcal{O}(\xi^2)$ and above in the following derivation of the state prepared by the entanglement swapping process with the caveat that this is justified only when the active modes into the swapping process are distinguishable.

The Hamiltonian governing the SFG process is
\begin{equation}
\hat{H}_{\text{SFG}}=\int_\mathcal{V}\text{d}\mathbf{r}\tilde{\chi}^{(2)}\hat{\mathbf{E}}_\text{a1}^{(+)}\hat{\mathbf{E}}_\text{a2}^{(+)}\hat{\mathbf{E}}^{(-)}_\text{SFG}+\hc,
\end{equation}
where $\tilde{\chi}^{(2)}$ is the second-order nonlinear susceptibility tensor, subscripts a1 and a2 refer to the active fields, respectively from sources 1 and 2, that are converted to the sum frequency field, $\hc$ is the Hermitian conjugate. The electric field operators are defined as
\begin{equation}
\hat{\mathbf{E}}_j(\mathbf{r},t) = \hat{\mathbf{E}}_j^{(+)}(\mathbf{r},t)+\hat{\mathbf{E}}_j^{(-)}(\mathbf{r},t),
\end{equation}
where $j$ indexes the active signal and idler fields and the SFG field over the subscripts $\{a1, a2, \text{SFG} \}$, and
\begin{align}
\hat{\mathbf{E}}_j^{(+)}(\mathbf{r},t)&=i\int_0^\infty  \frac{\text{d}\omega_j}{2 \pi} \hat{\mathbf{e}}_j\mathcal{E}_j (\mathbf{r}) e^{\ii [\mathbf{k}_j(\omega_j)\cdot \mathbf{r}- \omega_j t]} \hat{a}_j(\omega_j), \\
\hat{\mathbf{E}}_j^{(-)}(\mathbf{r},t)&=i\int_0^\infty  \frac{\text{d}\omega_j}{2 \pi} \hat{\mathbf{e}}_j^* \mathcal{E}_j^* (\mathbf{r}) e^{-\ii[\mathbf{k}_j(\omega_j)\cdot \mathbf{r}- \omega_j t]} \hat{a}_j^\dagger(\omega_j),
\end{align}
where $\hat{\mathbf{e}}_j$ is the unit polarization vector, $\mathbf{k}_j$ is the wavevector, $\omega_j$ the angular frequency, $\hat{a}_j^\dagger (\omega_j)$ and $\hat{a}_j (\omega_j)$ are respectively the creation and annihilations operators with commutator
\begin{equation}
[ \hat{a} (\omega_j) , \hat{a}^\dagger (\omega_j')] = 2 \pi \, \delta (\omega_j - \omega_j'),
\end{equation} 
and $\mathcal{E}_j (\mathbf{r}) =  \sqrt{\frac{\hslash \omega}{2 \epsilon_0 n_j (\omega_j) c}} u_j (\mathbf{r})$ is the single-photon electric field amplitude with material refractive index $n_j(\omega_j)$, speed of light in vacuum $c$, and waveguide mode $u_j(\mathbf{r})$. Similar approaches are detailed in \cite{Fiorentino2007,Milonni1995,Blow1990a}.

We select a crossed-polarization scheme and take $\tilde{\chi}^{(2)}$ to be the element from the full nonlinear tensor that couples the zyy crystallographic axes, which allows us to reduce the vector equations to a scalar problem \cite{Bierlein}. For simplicity we collect constant factors into $\chi^{(2)}$ in this theory section (the absence of the overtilde indicating the presence of the constants), but they are shown in detail in Appendix \ref{sec:CountRates}.
 With this, the Hamiltonian simplifies to
\begin{equation} \label{eq:Hamiltonian}
\begin{split}
&\hat{H}_{\text{SFG}}=\chi^{(2)}\int_\mathcal{V}\text{d}\mathbf{r}\int_0^\infty \text{d}\omega_{a1} \text{d}\omega_{a2}\text{d}\wsfg \\
&\left\{ \exp\big[ \ii \left( \mathbf{r}\cdot \Delta \mathbf{k} - \Delta \omega t \right) \big] \hat{a}_{a1}\hat{a}_{a2}\hat{a}^\dagger_\text{SFG}+\hc\right\},
\end{split}
\end{equation}
where
\begin{align}
&\Delta \mathbf{k} = \mathbf{k}_{a1}(\omega_{a1})+\mathbf{k}_{a2}(\omega_{a2})-\mathbf{k}_\text{SFG}(\wsfg) + \mathbf{k}_\Lambda, \\
&\Delta \omega = \omega_{a1}+\omega_{a2}-\wsfg,
\end{align}
with $\mathbf{k}_\Lambda$ accounting for the quasi-phase matching contribution. We have assumed that the field amplitudes are slowly varying in frequency and can be taken outside the integrals and then absorbed them into $\chi^{(2)}$. To first order, the state output after the SFG waveguide is described as 
\begin{equation}
\ket{\Psi(t)}_\text{out} \approx \ket{\Psi(t_0)}_\text{in}-\frac{\ii}{\hslash} \int_{t_0}^t\hat{H}_\text{SFG}(t')\, \text{d}t'\ket{\Psi(t_0)}_\text{in}.\label{eq:outputState}
\end{equation}
We select our SFG waveguide parameters such that SFG can only take place if a photon from each source is present, and take the input state to be
\begin{equation}
\begin{split}
\ket{\Psi (t_0)}_\text{in}=\int_0^\infty \text{d}\wione\text{d}\witwo\text{d}\wsone\text{d}\wstwo\Phi_1(\wione,\wsone)& \\
  \times \Phi_2(\witwo,\wstwo) \aione^\dagger\aitwo^\dagger\asone^\dagger\astwo^\dagger \vacuum &, \label{eq:inputState}
\end{split}
\end{equation}
where the frequency dependence of the creation operators has been suppressed and where $\Phi_j$ is the JSA of source $j \in \{1,2\}$. We assign the active and bystander fields as a1$\leftrightarrow$s1, b1$\leftrightarrow$i1, a2$\leftrightarrow$i2, and b2$\leftrightarrow$s2. Combining Eq.~\eqref{eq:outputState} and \eqref{eq:inputState}, extending the limits of the temporal integral to be from $-\infty$ to $\infty$, writing the phase-matching in a general form as
\begin{equation}
\Pi(\waone,\watwo,\wsfg) = \int_{0}^{L} \text{d}z \exp \left( -i \, \Delta \mathbf{k} \, z \right),
\end{equation}
suppressing the time-dependence in the states, and noting that the Hermitian conjugate term of the Hamiltonian acting on the input state gives zero, we find
\begin{equation}
\begin{split} \label{eq:threeFreqState}
&\ket{\Psi}_\text{out}= \ket{\Psi}_\text{in}-\frac{\ii \chi^{(2)}}{\hslash} 
\int_0^\infty \text{d}\wsfg\text{d}\wbone \text{d}\wbtwo\\
&\times \psi(\wbone,\wbtwo,\wsfg) \asfg^\dagger(\wsfg)\abone^\dagger(\wbone)\abtwo^\dagger(\wbtwo)\vacuum,
\end{split}
\end{equation}
where the three-frequency joint spectral amplitude is
\begin{equation} \label{eq:psi}
\begin{split}
\psi(\wbone,\wbtwo,\wsfg)= \int_0^\infty  \text{d}\watwo \text{d}\waone  \Pi(\waone, \watwo, \wsfg)& \\
\times \delta(\wsfg - \waone - \watwo) \Phi_1(\wbone,\waone)\Phi_2(\watwo,\wbtwo)&.
\end{split}
\end{equation}
Note that entanglement can only be swapped if there is entanglement in the input states to begin with, \ie $\Phi_1$ and $\Phi_2$ are both inseparable in their frequency arguments such that $\Phi(\wi,\ws) \neq F(\wi) G(\ws)$ where $F$ and $G$ are arbitrary functions that depend upon their arguments only. Armed with this three-photon state, our task is now to determine what phase matching function (SFG crystal parameters), and heralding measurement swap the input entanglement to generate the most desirable output entangled biphoton state.

\subsection{Heralding Swapped Entanglement}
It is worthwhile to consider the use of two categories of herald detectors. The first category detect the arrival of a photon without resolving its frequency and we refer to them as ``non-resolving'' detectors. The second category, ``frequency resolving'' detectors, report the frequency of the herald photon to within some resolution limit. A dispersive element can be combined with an array of non-resolving detectors, as shown in Fig.\ \ref{fig:setup}, to make a frequency resolving detector. Frequency non-resolving detectors offer simplicity and lower cost as advantages over frequency resolving detectors, so we start with consideration of entanglement swapping with frequency non-resolving detection.

Just as heralding one photon from an SPDC source with a frequency non-resolving detector can produce either a pure or a mixed single photon state depending on the separability of $\Phi(\wi,\ws)$ \cite{URen2005,Smith2009}, the purity of the output biphoton state after measurement of the SFG photon is set by the  separability characteristics of $\psi(\wbone,\wbtwo,\wsfg)$. A pure state biphoton is created only if $\psi$ can be factored such that
\begin{equation} \label{eq:separabilityCrit}
\psi(\wbone,\wbtwo,\wsfg)=P(\wbone,\wbtwo)Q(\wsfg),
\end{equation}
where $P$ is a function that is non-separable in $\wbone$ and $\wbtwo$ and $Q$ is a function of $\wsfg$ only. If frequency non-resolving heralding is performed on a $\psi$ that does not meet this separability criterion, then the output biphoton will be in an undesirable mixed state.

\subsubsection{Simple Model: Infinitely Long Crystals} \label{sec:toy}
The simplest model that achieves heralded entanglement swapping is perfect anticorrelated phase-matching in the SFG crystal. Let $k'(\omega) = \partial k (\omega)/\partial \omega = 1/ \nu_g$ be the inverse of the group velocity $\nu_g$, referred to as the group slowness. Anticorrelated phase-matching means
\begin{align}
k' (\bar{\omega}_{a1}) &= k' (\bar{\omega}_{a2}) \label{eq:anticorrelated} \\
k' (\bar{\omega}_{a1}) &\neq k' (\bar{\omega}_{\text{SFG}}) \\
\Pi = \delta(\waone& + \watwo - \bar{\omega}_{\text{SFG}}), \label{eq:toyPMdelta}
\end{align}
where $\bar{\omega}_i$ is the central frequency of $\omega_i$, around which $k(\omega_i)$ is Taylor expanded, and the Dirac delta function in Eq.\ \eqref{eq:toyPMdelta} results from assuming perfect phase-matching due to an infinitely long SFG crystal. Frequencies that satisfy this phase-matching condition are oriented along the difference-frequency axis, as illustrated in Fig.\ \ref{fig:sourceJSI}, from whence the term anticorrelated. With this phase-matching, $\psi$ becomes manifestly separable in $\wsfg$ with the form
\begin{equation}
\begin{split} \label{eq:toyEta}
\psi(\wbone,\wbtwo,\wsfg)= \delta (\bar{\omega}_{\text{SFG}} -\omega_{\text{SFG}}) \\
\times \int_0^\infty d\waone \Phi_1 (\wbone, \waone)  \Phi_2 (\bar{\omega}_{\text{SFG}} - \waone, \wbtwo).
\end{split}
\end{equation}
The integral enforces entanglement of the bystander modes, as can be seen by taking the source lengths $L \to \infty$ and assuming the source and SFG crystals are identical, (which implies $\bar{\omega}_{\text{SFG}} = \bar{\omega}_p$), yielding
\begin{equation} \label{eq:explicitDeltaPsi}
\psi (\wbone,\wbtwo,\wsfg) = \delta (\bar{\omega}_{p} -\omega_{\text{SFG}}) \delta(\wbone + \wbtwo - \bar{\omega}_{p}).
\end{equation}
Eq.\ \eqref{eq:explicitDeltaPsi} satisfies the separability criterion (Eq.\ \eqref{eq:separabilityCrit}), so using the SFG photon as a herald leaves the remaining signal/idler biphoton in an ideal state that is both pure and maximally entangled.

In contrast, perfect correlated phase matching (satisfied by frequencies oriented along the sum-frequency axis of Fig.\ \ref{fig:sourceJSI}) with 
\begin{align}
\left[ k'(\bar{\omega}_{\text{a}1}) - k'(\bar{\omega}_{\text{SFG}}) \right] &= -[k'(\bar{\omega}_{\text{a}2}) - k'(\bar{\omega}_{\text{SFG}})] \\
\Pi &= \delta \left( \waone - \bar{\omega}_{\text{a1}} - \watwo + \bar{\omega}_{\text{a2}}  \right)
\end{align}
in an infinitely long SFG crystal gives,
\begin{equation}
\begin{split}
\psi(\wbone, \wbtwo,\wsfg) = &\Phi_1 \left[ \wbone , \left(\wsfg - \Delta \bar{\omega}_a\right)/2 \right] \\
\times& \Phi_2 \left[ \left(\wsfg + \Delta \bar{\omega}_a \right)/2,\wbtwo \right],
\end{split}
\end{equation}
where $\Delta \bar{\omega}_a = \bar{\omega}_{\text{a1}} - \bar{\omega}_{\text{a2}}$. Taking $L \to \infty$,
\begin{equation}
\begin{split}
\psi(\wbone, \wbtwo,\wsfg) = & \delta \left( \frac{\wsfg - \Delta \bar{\omega}_a}{2}+\wbone- \bar{\omega}_p \right)  \\
\times& \delta \left( \frac{\wsfg + \Delta \bar{\omega}_a}{2}+\wbtwo - \bar{\omega}_p \right),
\end{split}
\end{equation}
and it is clear that $\wsfg$ is manifestly inseparable from both $\wbone$ and $\wbtwo$. Thus, with monochromatic pumps, long crystals, and heralding with the SFG photons directed to frequency non-resolving detectors, correlated phase-matching in the SFG crystal produces undesirable output states, while anticorrelated phase-matching heralds pure-state entangled biphotons.

This can be understood through the availability or erasure of frequency information. Correlated phase matching allows determination of $\waone$ and $\watwo$ through measurement of $\wsfg$, which simultaneously collapses the values of $\wbone$ and $\wbtwo$. Anticorrelated SFG erases information about the difference between the input frequencies, so measurement of $\wsfg$ does not allow determination of the input frequencies and the quantum superposition of the bystander modes is preserved.

No real experimental system will have perfectly delta-correlated phase matching, so it is necessary to consider mathematical tools for assessment of the effects of finite length and pulsed pump beams on the purity of the heralded biphoton state.

\subsubsection{Analytic Model: The Gaussian Phase-Matching Approximation}
To facilitate analytical investigation, we approximate both the pump pulses and the phase-matching functions as Gaussians, such that
\begin{align}
&\Pi(\waone, \watwo, \wsfg) \approx L \exp \left[- \frac{(L \, \Delta \widetilde{k} )^2 }{2 \sigma_\pi^2}\right] \label{eq:GaussPM} \\
&\Phi(\ws,\wi) \approx A \, L \exp \left[ -\frac{(\ws + \wi - \wpbar)^2}{2 \sigma_p^2} -\frac{(L \, \Delta \widetilde{k})^2}{2 \sigma_\pi^2} \right] \label{eq:GaussPhi}
\end{align}
where $A$ is the pump peak power, $\sigma_\pi$ is the Gaussian width (explained below), and we assume that the wavevectors are co-oriented along the waveguide ($z$) axis, allowing the use of a scalar $\Delta k$ with implicit frequency dependence. We neglect the phase factor in $\Pi(\waone, \watwo, \wsfg)$ as it is irrelevant to the analysis in this Section. Recall that $k(\omega) = n(\omega) \omega/c$. It is convenient for separability analysis to define $\Delta \widetilde{k} = c\, \Delta k$ and absorb the factor of $c$ into the definition of $\sigma_\pi = \kappa c/L$, where $\kappa=12.8831$ is the fit parameter that best matches a Gaussian width to the exact sinc functional form. The Gaussian approximation avoids the added complexity of evaluating the integrals of products of sinc functions, which are found in exact phase-matching models, (and are resolved numerically in section \ref{sec:simulation}), by allowing analytic integration of
\begin{widetext}
\begin{align} \label{eq:psiGaussian}
&\psi(\wbone,\wbtwo,\wsfg) = \int_0^\infty \diff \waone \, \Pi (\waone, \wsfg - \waone, \wsfg) \Phi_1 (\wbone, \waone) \Phi_2 (\wsfg - \waone, \wbtwo)  \\
&= L_{\text{SFG}} L^2 A^2 \int_0^\infty \diff \waone \, \exp \Big[ -\frac{\left(\wbone +\waone - \wpbar \right)^2+ \left(\wsfg - \wbone +\wbtwo - \wpbar \right)^2}{2 \sigma_p^2} -\frac{\left( \Delta \widetilde{k}_1\right)^2 + \left( \Delta \widetilde{k}_2\right)^2}{2 \sigma_\pi^2} - \frac{\left( \Delta \widetilde{k}_{\text{SFG}} \right)^2}{2 \sigma_{\text{SFG}}^2} \Big], \nonumber
\end{align}
\end{widetext}
where $\sigma_{\text{SFG}} = \kappa c/L_{\text{SFG}}$. Taking the refractive index variation over the wavelength range of each field to be small, we set $n_j = n(\omega_j) \approx n(\bar{\omega}_j)$, for $j \in \{p, s, i\}$. Eq. \eqref{eq:psiGaussian} satisfies the separability criterion (Eq.\ \eqref{eq:separabilityCrit}) when the prefactors for the cross-terms $\wsfg \wbone$ and $\wsfg \wbtwo$ are both zero. However, this occurs only when
\begin{equation}
-(n_p - n_s) (n_p - n_i) = \frac{\sigma_\pi^2}{\sigma_p^2},
\end{equation}
which \emph{is the condition for separable input states}, i.e.\ $\Phi(\ws,\wi) = F(\ws) G(\wi)$, which have no entanglement to be swapped. Thus, the frequency non-resolving measurement will produce mixed states for any realistic source that produces entangled output biphotons.

Why is separability in accordance with Eq.\ \eqref{eq:separabilityCrit} achievable in the infinite crystal limit, but not with finite crystals? Because using an infinitely long SFG crystal produces a monochromatic output field, which resolves $\wsfg$. This implies that we must use a frequency-resolving heralding scheme to achieve high-purity output biphoton states.

\subsubsection{Figures of Merit: Purity and Negativity}
In this Section, we develop a discretized description of the quantum state and review the mathematical machinery necessary for numerical simulation of a realistic model system. Consider heralding through frequency-resolving measurement of the SFG photon with outcomes forming a discrete set of disjoint frequency bins. We assume perfect detection with unit quantum efficiency, no dark counts, and lossless optical elements. Thus, herald detections occur uniquely after successful SFG (the second term in Eq.\ \eqref{eq:threeFreqState}), and always indicate the presence of a biphoton in the bystander fields.

Using a discretized frequency-bin description, the density matrix entries with $\wbone$ indexed by $\{j,j'\}$, $\wbtwo$ indexed by $\{k,k'\}$, and $\wsfg$ indexed by $\{l,l'\}$ are
\begin{equation} \label{eq:reducedDensityMatrix} 
\begin{split}
\rho(j,k,l,j',k',l')= & \\
\psi(\omega_j, \omega_k, \omega_l) \psi^*(&\omega_{j'}, \omega_{k'}, \omega_{l'}) \Delta \wbone \, \Delta \wbtwo \, \Delta \wsfg,
\end{split}
\end{equation}
where $\Delta \omega_\alpha$ refers to the spacing between the frequency grid points for field $\alpha \in \{\text{b1, b2, SFG}\}$, so $\rho$ values here are probabilities, not probability densities. The spacings must, of course, be set smaller than the scale of the smallest structures in $\psi$ in order to resolve those features, and it is useful to keep in mind the experimentally accessible spectroscopic resolution limit of the SFG field of about 25 GHz (0.16 rad/ps) \cite{Davis2016, Kuo2016}.

If the SFG photon is measured with a frequency non-resolving detector, then $\wsfg$ is traced out, yielding the reduced density matrix \cite{Kolenderski2009,Kolenderski2009b}
\begin{equation} \label{eq:reducedDensityMatrix2} 
\begin{split}
\rho_r(j,k,j',k')= \sum_l \psi(\omega_j, \omega_k, \omega_l) \psi^*(\omega_{j'}, \omega_{k'}, \omega_l)\\
\times \Delta \wbone \, \Delta \wbtwo \, \Delta \wsfg.
\end{split}
\end{equation}
Loss of information about the value of $\wsfg$ degrades the purity of the output state 
\begin{equation} \label{eq:purity}
\mathcal{P} = \text{Tr} (\rho^2 ),
\end{equation}
where Tr is the trace operation.

To describe heralding with a frequency-resolving detector, let the spectroscopic measurement of $\wsfg$ be described by a set of projective measurement operators $\{\hat{\Omega}_m \}$ with
\begin{align}
\hat{\Omega}_m &= \int_{\bar{\omega}_m-\Delta'/2}^{\bar{\omega}_m+\Delta'/2} \ket{\omega} \bra{\omega} \diff \omega, \label{eq:Omega} \\
\sum_{m=1}^N \hat{\Omega}_m &= \mathbb{I}, \label{eq:completeness}\\
\hat{\Omega}_m \hat{\Omega}_{m'} &= \delta_{mm'} \hat{\Omega}_m, \label{eq:orthonormality}
\end{align}
where $\ket{\omega} = \hat{a}^{\dagger} (\omega) \ket{\text{vac}}_{\text{SFG}}$ is a single photon in the SFG spatial mode with frequency $\omega$, $N$ is the number of frequency bins, $\bar{\omega}_m$ is the central frequency of the $m$th bin, $\Delta'$ is the frequency-space width of a measurement bin, $\mathbb{I}$ is the identity matrix, and $\delta_{mm'}$ is the Kronecker delta. In order for Eq. \eqref{eq:completeness} to hold, the range of frequencies measured $N \Delta'$ must exceed the range of frequencies produced via SFG . Otherwise, successful entanglement swapping events will go undetected. Conditioning on a herald detection alleviates the need for a vacuum outcome in Eq.\ \eqref{eq:completeness}, and Eq. \eqref{eq:completeness} and \eqref{eq:orthonormality} together imply that $\hat{\Omega}_m^\dagger \hat{\Omega}_m = \hat{\Omega}_m$. Detection of the frequency of the SFG photon projects the output state into
\begin{equation} \label{eq:RhoMeasured}
\hat{\rho}_m' (\wbone,\wbtwo,\wbone',\wbtwo') = \frac{\hat{\Omega}_m \hat{\rho} \,  \hat{\Omega}_m}{\text{Tr} \left( \hat{\Omega}_m \hat{\rho} \right)},
\end{equation}
with $\wsfg = \bar{\omega}_m$.

To account for the limited resolution of a realistic measurement, let $\Delta$ be the separation between the frequency grid values used for computation and enforce $\Delta < \Delta'$. If we posit that $\Delta'/\Delta = Q$, which we take to be a positive integer, then we can divide Eq.\ \eqref{eq:completeness} by grouping together the $Q$ operators that comprise the $n$th measurement outcome, starting at $n=1$, to make the operator for the resolution-limited measurement
\begin{equation}\label{eq:binDef}
\hat{\widetilde{\Omega}}_{n} = \sum_{m=(n-1)Q+1}^{nQ} \hat{\Omega}_{m}.
\end{equation}
Thus Eq.\ \eqref{eq:RhoMeasured} generalizes to a linear combination of measurement operators at the discretization size with the straightforward substitutions $\hat{\Omega}_m \to \hat{\widetilde{\Omega}}_{n}$ and $\hat{\rho}_m' \to \hat{\widetilde{\rho}}_n'$. The populations  $\hat{\widetilde{\rho}}_n' (\wbone,\wbtwo,\wbone,\wbtwo)$ compose the JSI of the biphoton prepared upon detection of the $n$th herald outcome. In principle, the upper bound on $n$ can be an arbitrarily large integer, but in practice this upper bound is resource constrained. The tradeoff between the number of grid points considered and the computational resources required to calculate the density matrix are discussed in Appendix \ref{sec:Simulation}. $Q$ must be chosen to give a good estimate for the purity, but computational resources requirements scale sharply with $Q$.

The entanglement in the state of the heralded bystander photons can be quantified using the negativity of the density matrix \cite{Vidal2002}
\begin{equation}
\mathcal{N}(\hat{\rho})\equiv (\norm{\hat{\rho}^{\Gamma_\textrm{i1}}}-1)/2, \label{eq:negDef}
\end{equation}
where ${\Gamma_\textrm{i1}}$ denotes the partial transpose operation with respect to subsystem i1 and $\norm{\hat{\rho}}\equiv\trace\left[(\hat{\rho}^\dagger\hat{\rho})^{1/2}\right]$. The presence of negative eigenvalues $\mu_i$ of the partially   transposed density matrix $\hat{\rho}^{\Gamma_\textrm{i1}}$ implies entanglement, and the negativity can also be expressed as the sum of these negative eigenvalues
\begin{equation}
\mathcal{N}(\hat{\rho}) = \sum_i \left| \mu_i \right| .
\end{equation}
The system is entangled in subsystem i1 if the negativity is positive. The negativity gives an upper bound on the amount of entanglement distillable from the state for teleportation \cite{Bouwmeester1997}. An investigation of the behavior of the negativity using simple density matrices is provided in Appendix \ref{app:negativity}. 

\section{Numerical Simulation} \label{sec:simulation}
What SFG crystal parameters optimize the entanglement swapping process and produce output biphotons with the highest negativity and purity? In this Section, we answer this question using numerical studies that include exact phase-matching functions. Let the SFG waveguide poling period $\Lambda_{\text{SFG}} = \Lambda$, so the central frequencies of the interacting fields $\{\wpbar, \bar{\omega}_s, \bar{\omega}_i \}$ are the same in all crystals, but allow the SFG waveguide length $L_{\text{SFG}}$ to be free to vary. We set the average pump power $P_{\text{avg}}$ such that the probability of a single photon-pair from a source is $|\xi|^2 = 0.1$, calculate $\psi$, use $\psi$ to calculate the herald count rate
\begin{equation}
R_H  = (2 \pi)^3 \Delta \wbone \, \Delta \wbtwo \, \Delta \wsfg   \sum_{j,k,l} |\psi(j,k,l)|^2 R_R,
\end{equation}
where $R_R$ is the pump laser repetition rate, and calculate both negativities and purities via the appropriate density matrices.

We model sources 1 and 2 in Fig.\ \ref{fig:setup} as periodically-poled potassium titanyl phosphate (KTP) crystal waveguide with length $L$ and a poling period $\Lambda=8.33\, \mu$m. The type-II phase matching function of the source is
\begin{align}
&\Pi_{\text{source}} =  L\, \text{sinc} \left(L \,\Delta k/2 \right) \text{Exp}\left( - i \, L \,\Delta k/2 \right), \label{eq:PMexact} \\
&\Delta k =   \frac{n_y (\omega_p) \omega_p}{c} - \frac{n_z (\omega_i) \omega_i}{c} - \frac{n_y (\ws) \ws}{c} - \frac{2 \pi}{\Lambda}, \label{eq:dkSource}
\end{align}
where $n_j (\omega)$ is the frequency-dependent refractive index along the crystallographic axis $j \in \{y, z\}$ \cite{Kato2002}, and the last term in Eq.\ \eqref{eq:dkSource} includes the first-order quasi-phase-matching effects of periodic poling \cite{Fejer1992}. We assume that that the only spatial mode excited in all frequency bands of all PPKTP waveguides is the fundamental. 

We set the length of the source crystals $L=0.5$ mm so the simulation can be carried out in a reasonable time and then match the pump bandwidth $\sigma_p$ to the approximate phase-matching bandwidth, $\sigma_\pi = \sigma_p =7.7245$ rad/ps. When pumped with a laser that has central wavelength $\bar{\lambda}_p=405.0$ nm ($\bar{\omega}_p = 4.651$ rad/fs), these sources create signal and idler fields at $\bar{\lambda}_s = 609.6$ nm ($\bar{\omega}_s = 3.090$ rad/fs) and $\bar{\lambda}_i = 1207$ nm ($\bar{\omega}_i = 1.561$ rad/fs) respectively. We chose our simulation parameters based on real pump laser systems \cite{Taccor} with the highest repetition rates $R_R$ where appropriate pump power is achievable. Appendix \ref{sec:Simulation} contains further simulation details and the parameters used herein are listed in Table \ref{tab:realisticParameters}.

\begin{figure}[htb]
\begin{center}
\includegraphics[width=0.9 \columnwidth]{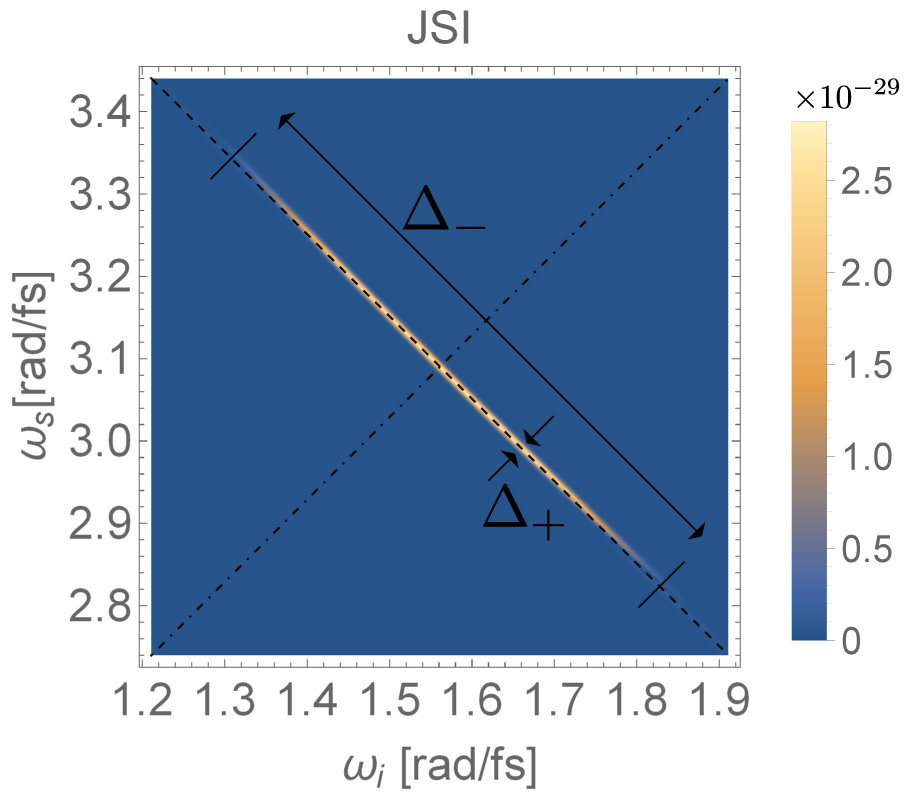}
\caption{$|\Phi(\wi,\ws)|^2$, the joint spectral intensity (JSI) from a single source with the phase matching function and parameters described in Sec.\ \ref{sec:simulation}. The two input sources are modeled to be identical. The dashed line is the difference-frequency axis and the dot-dashed line is the sum-frequency axis.}
\label{fig:sourceJSI}
\end{center}
\end{figure}

Assuming a Gaussian pump spectral profile
\begin{equation} \label{eq:PMbandwidth}
A(\omega_p) = \sqrt{\frac{P_{\text{avg}}}{\hslash \omega_p R_R \sigma_p \sqrt{\pi}}} \exp \left[ -\frac{(\omega_p - \bar{\omega}_p)^2}{2 \sigma_p^2} \right],
\end{equation}
with average power $P_{\text{avg}} = 1.380$ W and repetition rate $R_R = 1$ GHz, the joint spectral intensity (JSI, $|\Phi(\wi,\ws)|^2$) of a single source is shown in Fig.\ \ref{fig:sourceJSI}. The average power is set such that the probability of a photon pair being generated in a single source $|\xi|^2=0.1$, which sets the probability of the next-highest-order contribution to $|\xi|^4=0.01$.

Guided by insight from the simpler model of Section \ref{sec:toy}, we choose a pump wavelength and poling period to give an anticorrelated input JSA with a narrow width $\Delta_+$ along the sum-frequency axis (set by the pump bandwidth $\sigma_p$ and the phase-matching bandwidth $\sigma_\pi(L)$), and a broad width $\Delta_-$ along the difference-frequency axis (set by dispersion and  $\sigma_\pi(L)$). An extreme aspect ratio (e.g.\ $\Delta_-/\Delta_+ \gg 1$), which indicates a large number of time-frequency modes \cite{Nunn2013}, in addition to a small probability amplitude for the vacuum term in Eq.\ \eqref{eq:SPDCstate} are good heuristics for large negativity.

\begin{figure*}[htb]
\begin{center}
\includegraphics[width=0.99\textwidth]{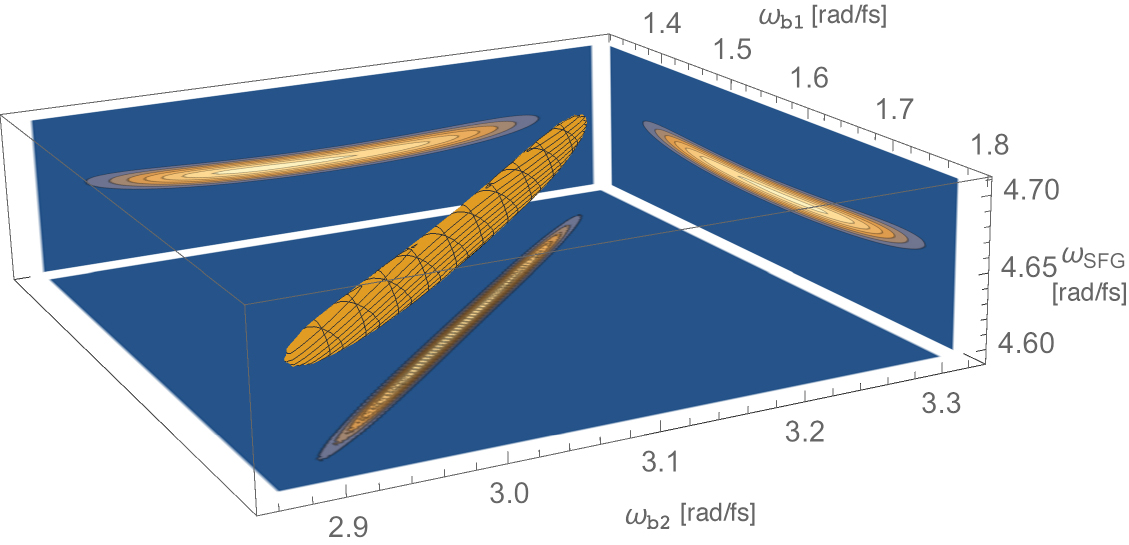}
\caption{3D Contour surface showing $|\psi(\wbone,\watwo,\wsfg)|^2$ with contour plots of the projections on the back planes. The 3D contour surface connects values of one-tenth of the maximum value of $|\psi|^2$. The tilt angle resulting from the correlation between $\wsfg$ and the sum frequency $\wbone+\wbtwo$, is more apparent in the output JSIs of Fig.\ \ref{fig:outputJSIs}.}
\label{fig:eta}
\end{center}
\end{figure*}

The SFG crystal has the same phase-matching function as given in Eq.\ \eqref{eq:PMexact}-\eqref{eq:dkSource} but with $L \to L_{\text{SFG}}$ and $\omega_p \to \wsfg$. Taking $L_{\text{SFG}} = L= 0.5$ mm, the three-frequency JSI $|\psi(\wbone,\wbtwo,\wsfg)|^2$ is visualized in Fig.\ \ref{fig:eta}. The expected rate at which heralded biphotons are produced with these lengths (whether or not $\wsfg$ is resolved) is $5.2\times10^{-3}$ biphotons/second, corresponding to one heralding event every 3.2 minutes.

Consider a spectroscopic measurement that resolves $\wsfg$ into 8 possible outcomes, indexed by $n$, each with frequency size $\Delta' = 3.862\times 10^{-3}$ rad/fs, i.e.\ 614.7 GHz. Let the probability-valued spectrum be denoted $p(\omega_{\text{SFG}}^n)$. To account for finite resolution, we use the incoherent sum prescribed by Eq.\ \eqref{eq:binDef} with $Q=3$ grid points. This choice underestimates the purity by a few percent due to discretization error, which means the stated purity values should be understood as lower bounds, but allows a full computational run to be carried out in a reasonable amount of time (see Appendix \ref{sec:Simulation}).

\begin{figure*}[tb]
\begin{center}
\subfigure[]{\includegraphics[width=0.48\textwidth]{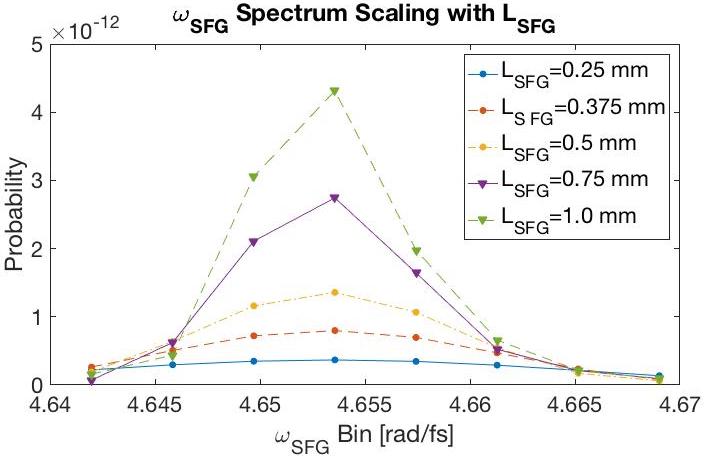} \label{fig:probability}}
\subfigure[]{\includegraphics[width=0.48\textwidth]{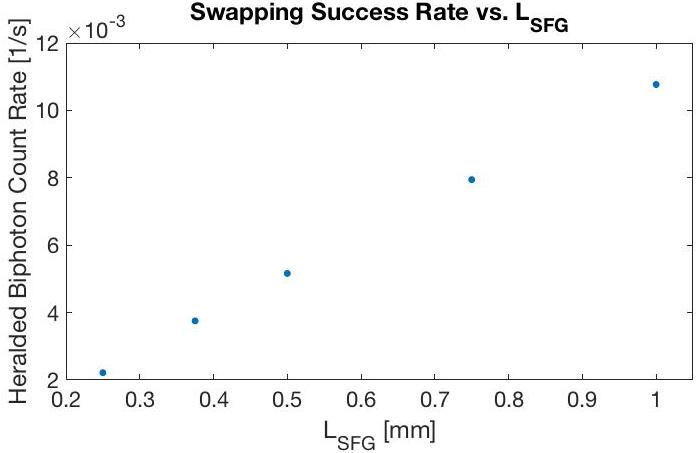} \label{fig:CountRatesVSLsfg}}
\caption{Scaling of the (a) $\wsfg$ spectrum $p(\omega_{\text{SFG}}^n)$, and the corresponding (b) rate of entanglement swapping events as $L_{\text{SFG}}$ varies. In (a) straight lines connect calculated points to serve as eye guides. Frequency-resolving measurement of $\wsfg$ gives discrete outcomes described after Eq. \eqref{eq:binDef} and labeled with the central frequency of the range of frequencies that constitute the corresponding frequency bin.}
\label{fig:spectrumAndRates}
\end{center}
\end{figure*}

\begin{figure*}[tb]
\begin{center}
\subfigure[]{\includegraphics[width=0.485\textwidth]{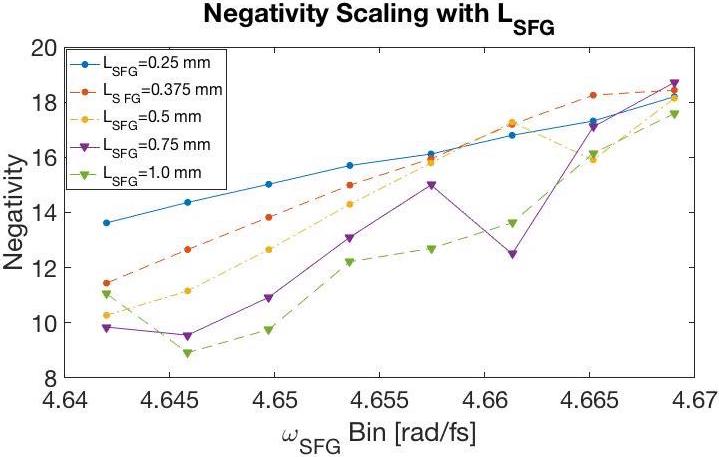} \label{fig:negativityResolved}} 
\subfigure[]{\includegraphics[width=0.49\textwidth]{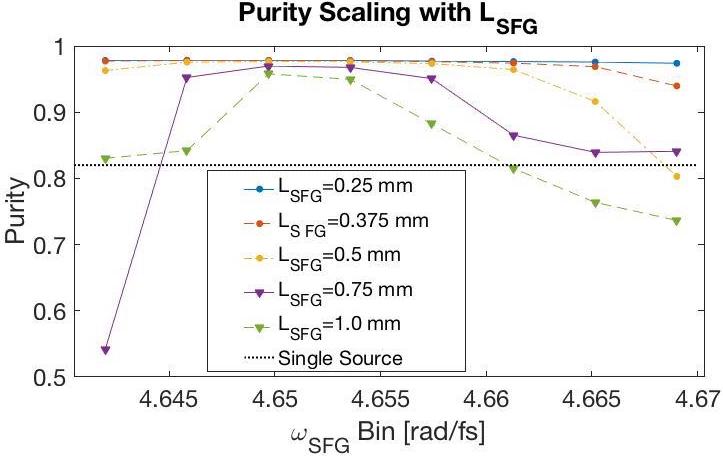} \label{fig:purityResolved}} 
\caption{Figures of merit with frequency-resolving detection. The (a) negativity $\mathcal{N}(\omega_{\text{SFG}}^n)$, and (b) purity $\mathcal{P}(\omega_{\text{SFG}}^n)$ for the output biphoton state prepared by each measurement outcome are shown for five $L_{\text{SFG}}$ values. A single SPDC source used in this experiment has a purity of 0.82 (indicated with the dashed reference line), and a negativity of 2.89 (not shown). The points indicate calculated values and are connected by lines to serve as eye guides, which should not be taken to represent the actual shape of the curve.}
\label{fig:ResolvedValues}
\end{center}
\end{figure*}

We now allow $L_{\text{SFG}}$ to vary while holding $L$ fixed at 0.5 mm. Fig.\ \ref{fig:spectrumAndRates} shows the $\wsfg$ spectrum and swapping success rate for many values of $L_{\text{SFG}}$. It is clear that $L_{\text{SFG}}$ sets the width and height of the $\wsfg$ spectrum, and thus the count rate. Fig.\ \ref{fig:ResolvedValues} displays the behavior of the negativity $\mathcal{N}$ and purity $\mathcal{P}$ w.r.t.\ the $\wsfg$ measurement outcome for several $L_{\text{SFG}}$ values. The broader frequency distributions output from short crystals (due to larger conversion bandwidths) correspond to more gradual changes in the state along the $\wsfg$ axis, which leads to higher purity. Broader distributions correspond to entanglement over more frequency modes and increased negativity. These improvements in negativity and purity trade-off with entanglement swapping success rates, which are higher with longer SFG crystals.

\begin{figure*}[p]
\begin{center}
\includegraphics[width=0.6\textwidth]{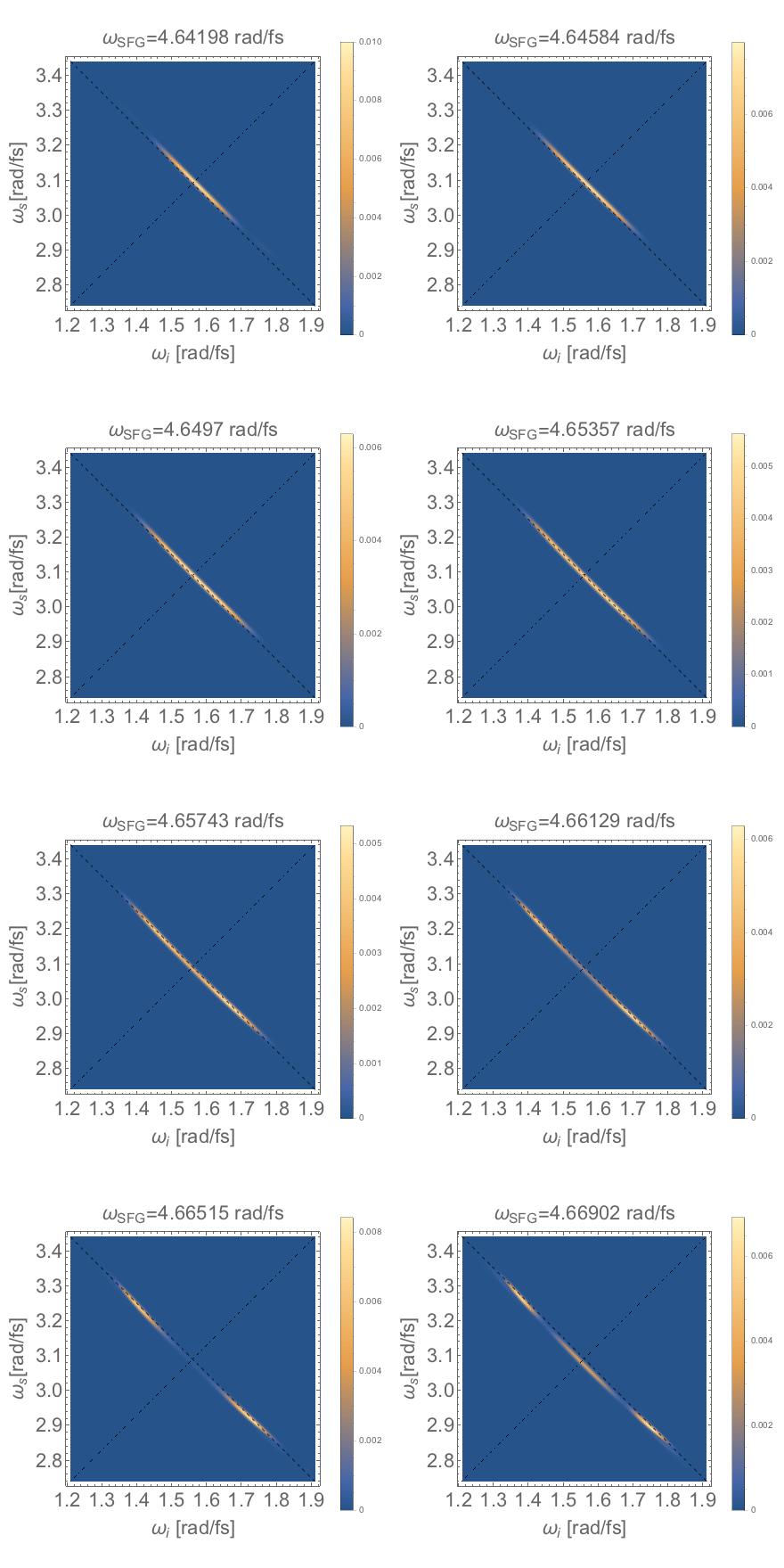}
\caption{Output JSIs with $L_{\text{SFG}} = 0.5$ mm. The dashed line is the difference-frequency axis and the dot-dashed line is the sum-frequency axis. Energy conservation requires that the sum frequency of the output biphoton decreases as the measured $\wsfg$ increases. The $\wsfg$ spectrum is relatively narrowband, so these differences are visible but small.}
\label{fig:outputJSIs}
\end{center}
\end{figure*}

Fig.\ \ref{fig:negativityResolved} shows a correlation between larger $\wsfg$ measurement values and larger negativity values because the distribution of the prepared biphoton is correspondingly broader along the difference-frequency axis. This can be clearly seen by comparing the JSIs for each $\wsfg$ measurement outcome, shown in Fig.\ \ref{fig:outputJSIs}. The slight shifts in the sum frequency of each JSI shifts with the $\wsfg$ measurement outcome so the total output photon energy is equal to the input photon energy.

It is illuminating to compare these entanglement-swapping-prepared biphoton states, which have negativities ranging from $9$ to $19$, to that of the biphoton state from a single SPDC source. Using our design parameters, negativity of the biphoton state from a single source is $2.89$. If the vacuum contribution were to be eliminated and the source produced only biphotons, then the negativity would increase to $28.9$ (a factor of $1/|\xi|^2$ increase in agreement with the linear relationship given in Eq.~\eqref{eq:toyNegativityScaling}). The vacuum contribution is removed by the swapping process, but the conversion process produces fewer entangled frequency modes within the resulting biphoton than are present in the source, so no outcome exceeds the ideal negativity of the source with the vacuum contribution removed (see Fig.\ \ref{fig:negativityResolved}). The net result is that probability of having a photon pair in a known time window is increased from 0.1 in the case of an a single SPDC source to near unity after detection of a herald signals successful swapping, which also substantially increases the negativity.

To compare the purity of biphoton states prepared through entanglement swapping to those prepared by a single SPDC source, it is important to note that the state of a biphoton generated through SPDC depends on the phase $\phi$ of the pump field used to create it. If the phase of the pump field is not resolved through measurement (and it is common practice to not resolve this phase), then the coherence elements between the biphoton and vacuum subspaces are lost \cite{Chou2005}. Appendix \ref{app:negativity} describes this in more detail. The purity of the biphoton state output from a single SPDC source in our design is 0.82. As shown in Fig.\ \ref{fig:purityResolved}, $L_{\text{SFG}}$ can be chosen such that all $\wsfg$ measurement values exceed 0.82, but as $L_{\text{SFG}}$ increases, outcomes with lower purity can occur.

\begin{figure*}[tb]
\begin{center}
\subfigure[]{\includegraphics[width=0.485\textwidth]{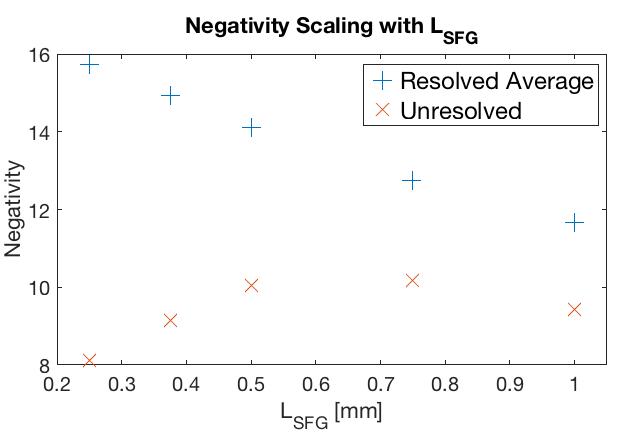} \label{fig:NegativityVSLsfg}}
\subfigure[]{\includegraphics[width=0.49\textwidth]{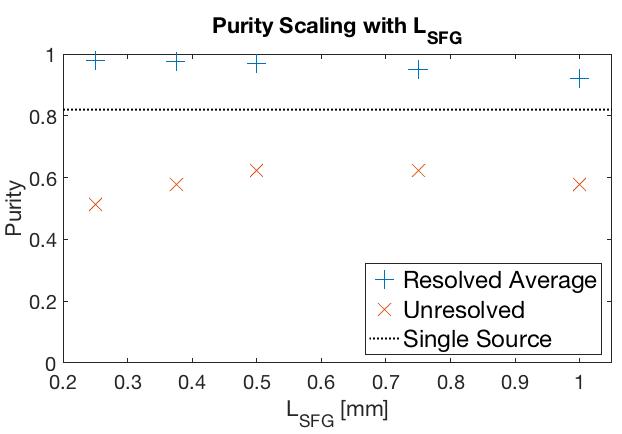} \label{fig:PurityVSLsfg}}
\caption{Comparison of the (a) negativity and (b) purity of output states prepared via either frequency resolving or frequency non-resolving heralding over five $L_{\text{SFG}}$ values. The effect of frequency resolution using the same measurement configuration as used in Fig.\ \ref{fig:ResolvedValues} and the weighted average of Eq.\ \eqref{eq:avgNegativity} is shown with the blue points, while the orange points show the values when the SFG photon is detected but $\wsfg$ is unresolved.}
\label{fig:valueScalingVLSsfg}
\end{center}
\end{figure*}

The purity and negativity will vary from shot to shot in accordance with the $\wsfg$ measurement outcome. A weighted average over the measurement outcomes
\begin{equation} \label{eq:avgNegativity}
\bar{\mathcal{A}} = \left( \sum_{n=1}^N p(\omega_{\text{SFG}}^n) \mathcal{A}_m \right) \Bigg/ \sum_{n=1}^N p(\omega_{\text{SFG}}^n),
\end{equation}
where $\mathcal{A}$ stands in for either the negativity $\mathcal{N}$ or the purity $\mathcal{P}$, gives the average values, i.e.\ expected performance, over many successful swapping events. If $\wsfg$ is not resolved, then the mixed output state gives negativity and purity values that are the same for each shot. Thus, the expected performance of entanglement swapping with frequency-resolving heralding can be compared to frequency non-resolving heralding by comparing the weighted averages for the frequency-resolving configurations to the unresolved values, as shown in Fig.\ \ref{fig:valueScalingVLSsfg} for many $L_{\text{SFG}}$ values. Frequency-resolving heralding clearly yields superior performance for both purity and negativity.

The negativity of all configurations in Fig.\ \ref{fig:NegativityVSLsfg} exceed the negativity of the biphoton state output from a single source (2.89), so entanglement swapping purifies (in the entanglement sense) the output state. In contrast, Fig.\ \ref{fig:PurityVSLsfg} shows that frequency non-resolving heralding offers inferior purity compared to a single source for all $L_{\text{SFG}}$ choices. Even though some $\wsfg$ measurement outcomes give biphoton sates with purity below that produced by a single source (see Fig.\ \ref{fig:purityResolved}), the average purity is improved for all frequency-resolving heralding configurations here considered. These averages could be further improved by rejecting heralding events that produce biphoton states with undesirable properties, at the cost of production rate.

\section{Discussion}
While entanglement swapping with frequency-resolving heralding produces entangled biphotons with superior negativity and purity compared to SPDC sources, the count rates are inferior by roughly 7 orders of magnitude. This is compensated in a sense by creating, through heralding, near-unity probability of having an entangled pair in a known time window. While use of high repetition rate pump lasers can improve the count rate, the low probability of success during any given pump pulse remains a substantial challenge. Count rates can be modestly increased through increasing the source crystal lengths $L$ and maximimizing $L_{\text{SFG}}$ as allowed by negativity and purity tolerances, but keeping the sources in the low-gain regime with $|\xi|^2 \approx 0.1$ limits these improvements. Additionally the computational resources required to perform the simulation scale sharply with the source lengths $L$.
 
Increasing the pump repetition rate $R_R$ to limits set by technical constraints such as achievable pump power and detector recovery times will increase the rate of successful swapping (Eq.\ \eqref{eq:swapRate}). However, this seems like a less promising approach compared to increasing the single shot success probability while avoiding the false heralding problem due to indistinguishable inputs discussed in the introduction of Section \ref{sec:simulation}. Thus, investigation of whether four-wave mixing can offer better count rates through higher effective nonlinearity is a natural extension for future research.  Investigation of entanglement swapping outside the low-gain regime, where higher-order Hamiltonian terms that we neglect in Eq.\ \eqref{eq:outputState} contribute, may offer higher count rates and the heralded preparation of higher number states.

In summary, we propose a design for a source of heralded time-frequency-entangled photon pairs with high-dimensional frequency-bin encoding and give a detailed description of the mathematical machinery necessary for its characterization. Heralding with frequency-resolving detection produces high-purity ($\mathcal{P} \approx 0.97$) output biphoton states and improves the negativity compared to an SPDC source by roughly a factor of 5 (depending on the exact configuration chosen). The length of the SFG crystal and the size and number of detection bins sets the negativity, purity, and rate of successful entanglement swapping. Shorter SFG crystals offer superior negativity and purity due to broader conversion bandwidth, but inferior count rates, so a balance must be struck for a particular application.

The quantum erasure of which-path information via SFG that is essential for entanglement swapping has recently been identified as a key resource for quantum illumination and entanglement-enhanced metrology \cite{Lloyd08,zhuang17a, zhuang17b}, and the detailed calculations and design considerations we present are pertinent to implementation of those schema. One promising extension of this work is to produce output biphotons with the same central frequencies for all $\wsfg$ outcomes by deterministically frequency shifting the photon frequencies output from entanglement swapping depending on which SFG bin is detected \cite{Wright2017}. Another is measurement of photons produced with this apparatus in a pulse-mode (``temporal mode'') basis \cite{Reddy2015} would increase experimental complexity, but may offer even better negativity and purity values due to what appears to be a more natural basis choice. With improved count rates, this scheme would offer an ideal source for distributing entanglement resource over a network.

\begin{acknowledgements}
We thank K.\ Banaszek, D.\ Reddy, C.\ Jackson, A.~J.\ Vieta, and S.\ van Enk for helpful discussions, and A.\ Tamimi and A.\ H.\ Marcus for computer time. This material is based upon work supported by the National Science Foundation (NSF) under Grant No. 1620822, and by a Major Research Instrumentation grant from the NSF Office of Cyber Infrastructure, Principal Investigator Allen Malony, ``MRI-R2: Acquisition of an Applied Computational Instrument for Scientific Synthesis (ACISS),'' Grant no.: OCI-0960354. MGR was supported by NSF Grants QIS PHY-1521466 and AMO PHY-1406354. This work was partially funded by the HOMING programme of the Foundation for Polish Science (project no. Homing/2016-1/4) co-financed by the European Union under the European Regional Development Fund.
\end{acknowledgements}

\appendix
\section{Count Rates} \label{sec:CountRates}
In this Appendix, we include all constants in our calculations and predict photon pair creation rates for each source (cf.\ \cite{Fiorentino2007}), and for the entire entanglement swapping process. The sum-frequency generation (SFG) interaction Hamiltonian and output state for the single-biphoton subspace from an SPDC source in the low-gain regime are
\begin{align} 
&\hat{H}_{\text{SFG}} = \frac{\epsilon_0}{2} \int \diff^3 \mathbf{r} \, d \,\hat{E}_{\text{SFG}}^{(-)} \hat{E}_{\text{s1}}^{(+)} \hat{E}_{\text{i2}}^{(+)} + \text{H.c.} \\
&\ket{\psi}_{k} = \int_0^\infty  \diff \omega_{\text{s}k} \diff \omega_{\text{i}k} \Phi_k (\omega_{\text{i}k}, \omega_{\text{s}k}) \hat{a}_{\text{s}k}^\dagger (\omega_{\text{s}k}) \hat{a}_{\text{i}k}^\dagger (\omega_{\text{i}k}) \ket{\text{vac}}, \label{eq:singleSource} \\
&\Phi_k ( \omega_{\text{i}k}, \omega_{\text{s}k}) = b d \frac{(2 \pi)^2}{\sqrt{A_I}} \ell (\omega_{\text{s}k} + \omega_{\text{i}k}) \ell (\omega_{\text{s}k}) \ell (\omega_{\text{i}k})  \nonumber \\
& \quad \quad \times \Pi (\omega_{\text{s}k} + \omega_{\text{i}k}, \omega_{\text{s}k}, \omega_{\text{i}k}) \alpha (\omega_{\text{s}k} + \omega_{\text{i}k})
\end{align}
where $\epsilon_0$ is the permittivity of free space, $\hat{E}_j^{(\pm)}$ are the positive and negative frequency components of the quantized electric field operators, $\text{H.c.}$ is the Hermitian conjugate, $k \in \{1,2\}$ indexes the source, $\ws$ is the angular frequency of the signal photon, $\omega_i$ is the angular frequency of the idler photon, $\omega_p$ is the angular frequency of the pump and energy conservation $\omega_p = \ws + \wi$ is strictly enforced. Here $d$ is the effective nonlinear coefficient set by the material, $\ell (\omega)$ is the electric field per photon, $A_I$  is the effective area of the interaction (defined below) which depends on the transverse spatial distributions $u_j (x,y)$ of the interacting fields $j \in \{p, s, i \}$, $\Pi (\omega_p, \ws, \wi)$ is the phase matching function of the medium, and $\alpha (\omega_p)$ is the spectral pump pulse profile (assumed to have a Gaussian distribution).

Refining those definitions:
\begin{align}
b &=\frac{\epsilon_0}{2 \hslash (2 \pi)^3} \label{eq:b} \\
\ell (\omega) &= \sqrt{\frac{\hslash \omega}{2 \epsilon_0 n (\omega) c}}\\
\Pi(\omega_p, \ws, \wi) &= L \, \sinc \left[ L \Delta k (\omega_p, \ws, \wi) / 2 \right]  \nonumber \\
& \times \text{Exp}[-i\, L \Delta k (\omega_p, \ws, \wi) / 2]  \\
A_I = 1\Big / \Big( &\int_{-\infty}^\infty \diff x \diff y \,u_p (x,y) u_i^* (x,y) u_s^* (x,y) \Big)^{2} \\
\Delta k (\omega_p, \ws, \wi) &= k(\ws) + k(\wi) - k(\omega_p)+ \frac{q 2 \pi}{\Lambda} \\
\alpha (\omega_p) &= \sqrt{\frac{P_{\text{ave}}}{\hslash \omega_p \sigma_p \sqrt{\pi} R_R}} \exp \left[- \frac{(\omega_p - \bar{\omega}_p)^2}{2 \sigma_p^2} \right]
\end{align}
where $n(\omega)$ is the refractive index experienced by the photon in the source along the relevant polarization axis, $L$ is the longitudinal length of the nonlinear material that constitutes the source, $\sinc(x) = \sin(x)/x$ with $\sinc(0)=1$, $\Delta k$ is the momentum mismatch along the waveguide ($z$) axis, $q$ is the order of the quasi-phase matching and we use $q=1$. Additionally, $\Lambda$ is the crystal poling period, $\hslash$ is Planck's constant divided by $2\pi$, average pump power $P_{\text{ave}}$, repetition rate $R_R$, $\bar{\omega}_p$ is the central frequency of the pump, and $\sigma_p$ is the spectral bandwidth of the pump pulse.

Our design produces output photons with non-degenerate frequencies, and we use signal (idler) to refer to the higher (lower) frequency photon in accordance with historical convention. Given a pump pulse, the probability of photon-pair creation from a single source (with the source index $k$ suppressed) is given by
\begin{widetext}
\begin{align}\label{eq:sourceProbExact}
|\xi|^2 = \braket{\psi | \psi} = \frac{(2 \pi)^4 b^2 d^2}{A_I} &\int_0^\infty  \diff \ws \diff \wi \,\ell^2 (\ws + \wi) \ell^2 (\ws ) \ell^2 (\wi) |\alpha (\ws + \wi)|^2 |\Pi (\ws + \wi,\ws, \wi)|^2 \\
= \frac{ d^2 L^2 P_{\text{ave}}}{2^7 (2 \pi)^2 c^3 \sqrt{\pi} \epsilon_0 R_R \sigma_p A_I} &\int_0^\infty  \diff \ws \diff \wi  \frac{1}{n_y(\ws+\wi) n_y(\ws) n_z(\wi)}\exp \left[ - \frac{(\ws + \wi - \bar{\omega}_p)^2}{\sigma_p^2} \right] \times \nonumber \\ 
& \, {\sinc}^2 \left[ \frac{L \Delta k (\ws + \wi, \ws, \wi)}{2} \right].
\end{align}
\end{widetext}

Each SPDC source creates an output state of the form given in Eq. \eqref{eq:SPDCstate}, so the full input is the product state $\ket{\Psi}_1 \ket{\Psi}_2$. The annihilation operators in the SFG Hamiltonian given in Eq.\ \eqref{eq:Hamiltonian} remove contributions where only a single source produces photons. The lowest-order term of the input state that contributes to the SFG output state, using the configuration as shown in Fig.\ \ref{fig:setup}, is 
\begin{align} \label{eq:inputStateA}
\ket{\Psi}_{\text{in}} = &\int_0^\infty   \diff \waone \diff \watwo \diff \wbone \diff \wbtwo \, \Phi_1 (\wbone, \waone) \Phi_2 (\watwo, \wbtwo) \nonumber \\
& \cdot \hat{a}_{\text{a1}}^\dagger (\waone) \hat{a}_{\text{a2}}^\dagger (\watwo) \hat{b}_{\text{b1}}^\dagger (\wbone) \hat{b}_{\text{b2}}^\dagger (\wbtwo) \ket{\text{vac}} ,
\end{align}
where the argument ordering in the $\Phi$ functions indicates that the active photons directed into the SFG element are the signal from source 1 and the idler from source 2. We discuss the effects of higher-order terms at the end of this Appendix.

Eq.\ \eqref{eq:inputStateA} assumes that the pumps for each source are phase synchronized, as the pair creation process gets a phase imprint from the pump and if the pump lasers are not phase synchronized, phase diffusion introduces a relative phase shift between the photons sent into the SFG element. This phase difference will not affect conversion efficiency, but the output biphoton component of the state will acquire a phase shift that varies from shot to shot. Locking the pump phases to each other solves this problem.

Eq.\ \eqref{eq:threeFreqState}, representing the state output from the SFG crystal, can be rewritten as 
\begin{equation} \label{eq:threeFreqStateShort}
\ket{\Psi}_\text{out}= \ket{\Psi}_\text{in}-\ket{\psi}_{\text{SFG}},
\end{equation}
where $\ket{\psi}_{\text{SFG}}$ is the state produced by successful SFG. The corresponding probability of successful SFG is 
\begin{align} 
&|\Xi|^2 = \braket{\psi | \psi}_{\text{SFG}} \nonumber \\
= (2 \pi)^3 & \int_0^\infty \diff \wsfg \diff \wbone \diff \wbtwo \, |\psi (\wbone,\wbtwo,\wsfg)|^2, \label{eq:Psfg}
\end{align}
where
\begin{widetext}
\begin{equation} \label{eq:psiFull}
\begin{split}
\psi(\wbone,\wbtwo,\wsfg) = b d \frac{(2 \pi)^3}{\sqrt{A_I}}\int_0^\infty \diff \watwo \, \ell(\wsfg) \ell(\wsfg - \watwo) \ell(\watwo) \Pi(\wsfg-\watwo, \watwo, \wsfg)\\ \times \Phi_1(\wbone,\wsfg-\watwo) \Phi_2(\watwo,\wbtwo).
\end{split}
\end{equation}
\end{widetext}
The count rate for successful swapping is then
\begin{equation} \label{eq:swapRate}
R_H = |\Xi|^2 R_R.
\end{equation}

The higher-order contributions to the input state where both sources generate photon pairs and at least one source generates more than one photon pair are selected against by the fail detector. The most likely contribution to this is two pairs generated in one source and a single pair generated in the other, which triggers the fail detector whether or not an SFG photon is generated from two of the three input photons. In the case where both sources produce the same number of photon pairs and that number is greater than one, it is  possible for all input photons to be converted to SFG photons, which would not be detected by the fail detector, but is likely to trigger two simultaneous detections in the spectrometer. Though this contribution is negligible in our configuration as when $L_{\text{SFG}} = L = 0.5$ mm, the rate of false events from the leading-order term is $|\Xi|^4 R_R = 4.10 \times 10^{-14}$ events/sec, using SFG media with higher effective nonlinearities could result in this contribution being non-negligible. The use of a herald detection system that resolves the number of SFG photons in each frequency bin protects against this pitfall.

The ability of the fail detector to suppress multi-photon contributions is limited by its quantum efficiency $\gamma$. Given a herald detection, the leading-order probability that an extra photon-pair was generated in a source, but the remaining active photon was not detected by the fail detector is
\begin{equation}
P^{\text{swap}}_{\text{multi}} = (1 - \gamma) |\xi|^2.
\end{equation}
A superconducting nanowire single photon detector with $\gamma \approx 0.9$ and our scheme's $|\xi|^2 = 0.1$ gives $P^{\text{swap}}_{\text{multi}} = 0.01$, which means the output state generated after a herald detection has approximately 99\% entangled biphoton probability and 1\% multi-photon probability. In comparison, a single SPDC source generates a state with approximate probabilities of 10\% entangled biphoton, 1\% multi-photon, and 89\% vacuum. Larger quantum efficiency is better, but even in the worst case limit where $\gamma \to 0$, the state prepared by our entanglement swapping scheme has 90\% entangled biphoton probability, which is a substantial improvement over the SPDC state.

\section{Numerical Simulation Details} \label{sec:Simulation} 
We use Mathematica \cite{Mathematica} and Matlab \cite{Matlab} with the QETLAB toolbox \cite{QETLAB} to perform the numerical calculations presented in this paper. Table \ref{tab:realisticParameters} gives the parameters used in our presented calculations. $\psi$ is calculated from Eq.\ \eqref{eq:psiFull} with the integral evaluated numerically. The {\tt integrationPoints} parameter in Table \ref{tab:realisticParameters} is how many points are used with a trapezoidal rule method to perform this numerical integration.

The number of entries in the full three-frequency density matrix scales quickly with the number of signal, idler, and SFG frequency grid elements as
\begin{equation}
N_{\text{total}}= (N_s \times N_i \times N_{\text{SFG}})^2.
\end{equation}
We make efficient use of computational resources with utilization of sparse matrices, vectorized code, and direct calculation of the post-frequency-measurement density matrix of Eq.\ \eqref{eq:RhoMeasured}, which contains only
\begin{equation} \label{eq:nMeasured}
N_{\text{measured}}= N_{\text{SFG}} \times (N_s \times N_i)^2
\end{equation}
elements. Longer crystal lengths ($L,L_{\text{SFG}}$) correspond to narrower phase-matching bandwidths, which in turn requires finer frequency grid spacing. The negativity must be calculated in the signal/idler basis and entangled JSIs are oriented along diagonals, so this finer spacing requires more points in both the signal and idler directions. Memory requirements scale steeply with resolution improvement, \emph{e.g.\ }doubling the number of frequency grid points for both the signal and idler frequency grids requires 16 times more total memory. Thus, longer crystal lengths require substantially more memory and processor time for computation of purities and negativities. We choose the parameters in Table \ref{tab:realisticParameters} as realistic parameters that allow for calculations that complete in a reasonable amount of time. A full computational run for all data presented herein completes in 40 hours on a workstation with two 3.06 GHz, 6 core Xeon processors and 96 GB of RAM.

\begin{table}[t] 
\caption{Parameters used in the presented simulation.}
\begin{ruledtabular}
\begin{tabular}{ll}
\textbf{Parameter} &\textbf{Value} \\ \hline
Source length, $L$ & 0.50 mm  \\
Central pump frequency, $\bar{\omega}_p$ & 4.651 rad/fs  \\ 
Central signal frequency, $\bar{\omega}_s$ & 3.090 rad/fs  \\ 
Central idler frequency, $\bar{\omega}_i$ & 1.561 rad/fs  \\ 
Pump Gaussian bandwidth, $\sigma_p$ & 7.725 rad/ps  \\ 
SFG Frequency Spacing, $\Delta \wsfg$ & 1.287 rad/ps \\
Frequency Spacings, $\Delta \ws = \Delta \omega_i$ & 4.544 rad/ps \\ 
Poling Periods, $\Lambda=\Lambda_{\text{SFG}}$ & 8.33 $\mu$m  \\ 
Average pump power, $\Pavg$ & 1.380 W \\ 
Pump repetition rate, $R_R$ & 1.0 GHz \\ 
Nonlinear parameter, $d_{24}$ & 3.92 pm/V\\ 
Effective nonlinearity, $d$ & $2 d_{24}/\pi$ pm/V\\
Effective area, $A_I$ & 15 $\mu\textrm{m}^2$ \\ 
Grid points per outcome, $Q$ &  3 \\ 
{\tt integrationPoints} & 300 \\
\end{tabular}
\end{ruledtabular}
\label{tab:realisticParameters}
\end{table}
The step size for bystander frequencies is $\Delta \ws=\Delta \omega_i=\sigma_p / 1.7$, while $\Delta \omega_\text{SFG} = \sigma_p / 6$ is set for finer resolution. The SFG spectroscopic pixel bin size is set to $\Delta' = 3.862$ rad/ps (614.7 GHz), and includes $Q=3$ points from the underlying SFG frequency grid. $d_{24}$ is the nonlinear parameter for type-II phase-matching with the pump and signal polarized along the crystallographic y-axis and the idler polarized along the crystallographic z-axis. The pump bandwidth is set to match the phase-matching bandwidth for $L_{\text{SFG}}=L=0.5$ mm, which corresponds to $\sigma_p = \sigma_\pi = 7.725$ rad/ps.

\section{Negativity Behavior} \label{app:negativity}
In this Appendix, we investigate the behavior of the negativity as we adjust a simple model to give a sense of how it behaves. The four standard entangled two-qubit Bell states, \emph{e.g.} ($\ket{00} - \ket{11})/\sqrt{2}$, all have negativity 1/2 \cite{Rangamani2014}.

To investigate the behavior of the negativity for higher-dimensional quantum information encoding, we model the state out of a photon-pair source that directs one photon each to two parties, Alice ($A$) and Bob ($B$) with respective photon creation operators $\hat{a}^\dagger$ and $\hat{b}^\dagger$, as
\begin{equation}\label{eq:toyState}
\ket{\psi} = \sqrt{1-\eta} \ket{\text{vac}} + \sqrt{\eta} e^{i \phi} \sum^N_{j,k} \Psi_{j,k} \hat{a}^\dagger_{j} \hat{b}^\dagger_{k}    \ket{\text{vac}},
\end{equation}
where $\phi$ is the pump phase, $j$ and $k$ are frequency mode labels with integer values in the range $[1,N]$, and $\Psi_{j,k}$ is the complex-valued discretized joint spectral amplitude. For simplicity, we use a two-level model, which is good for approximating an SPDC source when $\eta \ll 1$. For convenience, we also introduce the shorthand notation
\begin{equation}
\ket{1}_A \ket{1}_B = \sum^N_{j,k} \Psi_{j,k} \hat{a}^\dagger_{j} \hat{b}^\dagger_{k} \ket{\text{vac}},
\end{equation}
where the frequency labels have been suppressed on the left-hand side of the equality.

\begin{figure}[htb]
\begin{center}
\includegraphics[width=0.98\columnwidth]{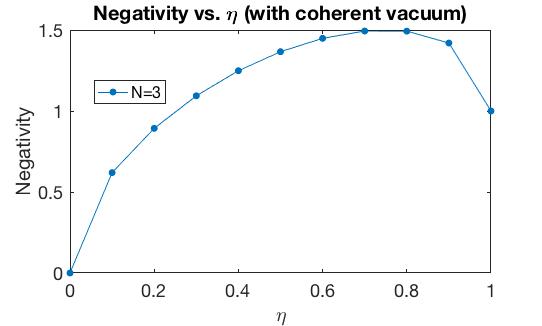}
\caption{Negativity as a function of pair-production efficiency $\eta$. Coherences between the vacuum and the biphoton subsystem are included.  Straight lines connect calculated points to serve as an eye guide.}
\label{fig:toyCoherent}
\end{center}
\end{figure}

As pointed out in the supplementary material of Chou \emph{et al.}, photon-pair creation processes such as SPDC are sensitive to the phase of the pump used to generate them \cite{Chou2005}. Naively creating the density matrix $\rho = \ket{\psi} \bra{\psi}$ includes the coherence terms $\ket{1}_A \ket{1}_B \bra{\text{vac}}_{AB} + \ket{\text{vac}}_{AB} \bra{1}_A \bra{1}_B$. In this case the vacuum could be used to coherently transfer quantum information, and plotting the negativity as a function of $\eta$ (see Fig.\ \ref{fig:toyCoherent}) shows a turn-around-point where the negativity decreases with increasing $\eta$. As $\eta \to 1$, the vacuum mode probability goes to zero. This reduction in the number of excited modes offers an intuitive explanation for the turn-around behavior.

\begin{figure*}[htb]
\begin{center}
\includegraphics[width=\textwidth]{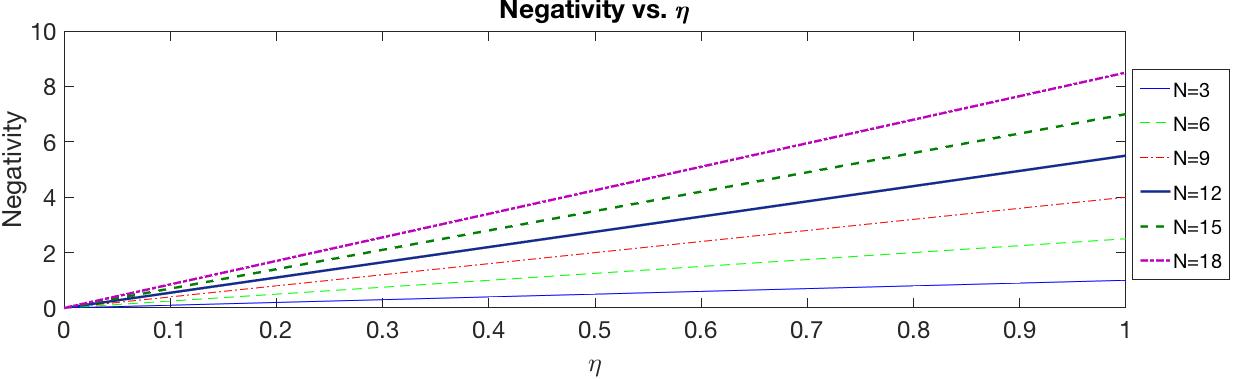}
\caption{Negativity $\mathcal{N}(\rho)$ vs. $\eta$ for multiple numbers of frequency modes, $N$. The input density matrices for this calculation have no coherence between the vacuum and biphoton subsystems.}
\label{fig:toyNmode}
\end{center}
\end{figure*}

In a real system, this coherence is preserved if the phase of the pump is measured, but decays at very fast optical frequencies otherwise. If the pump phase is not resolved, $\phi$ is traced out and  the coherences between the vacuum and one-pair subsystem vanish, yielding
\begin{equation} \label{eq:rhoCombo}
\rho = (1-\eta) \ket{\text{vac}}_{AB} \bra{\text{vac}}_{AB} + \eta \ket{1}_A \ket{1}_B  \bra{1}_A \bra{1}_B.
\end{equation}
Using a maximally-entangled frequency-anticorrelated state,
\begin{equation}
\Psi_{j,k} = \delta_{j,N+1-k}/\sqrt{N},
\end{equation}
where $\delta_{a,b}$ is the Kronecker delta, the negativity of the incoherent combination of Eq.\ \eqref{eq:rhoCombo} is shown in Fig.\ \ref{fig:toyNmode} for many values of the number of modes $N$, and follows the simple expression
\begin{equation} \label{eq:toyNegativityScaling}
\mathcal{N} = \frac{N-1}{2} \eta,
\end{equation}
which agrees with the Bell state negativity for $N=2$ and $\eta=1$. Thus, if quantum information is encoded in frequency bins, the number of bins chosen will influence the negativity. The negativity is not an intrinsic property of the continuous-variable state prepared by the SFG conversion process, but depends on the discretization imposed in detection of the herald and biphoton.

\phantom{t}


\begin{thebibliography}{99}

\bibitem{Gisin2007} N.~Gisin and R.~Thew, ``Quantum communication,'' Nat. Photonics {\bf1}, 165 (2007).

\bibitem{Torres2011} J.~P.~Torres, K.~Banaszek, and I.~A.~Walmsley, ``Engineering Nonlinear Optic Sources of Photonic Entanglement,'' Prog. Opt. {\bf 56}, 227 (2011).

\bibitem{Takeuchi2013} S.~Takeuchi, ``Recent progress in single-photon and entangled-photon generation and applications,'' Jpn. J. Appl. Phys. {\bf 53}, 030101 (2014). 

\bibitem{Barz2010} S.~Barz, G.~Cronberg, A.~Zeilinger, and P.~Walther, ``Heralded generation of entangled photon pairs,'' Nat. Photon. {\bf 4}, 553 (2010).

\bibitem{Yurke1992} B.~Yurke and D.~Stoler, ``Bell's-inequality experiments using independent-particle sources,'' Phys. Rev. A {\bf 46}, 2229 (1992).

\bibitem{Zukowski1993} M.~Zukowski, A.~Zeilinger, M.~A.~Horne, and A.~K.~Ekert, `` `Event-Ready-Detectors,' Bell Experiment via Entanglement Swapping,'' Phys. Rev. Lett. {\bf 71}, 4287 (1993).

\bibitem{Jennewein2001} T.~Jennewein, G.~Weihs, J.-W.~Pan, and A.~Zeilinger, ``Experimental Nonlocality Proof of Quantum Teleportation and Entanglement Swapping,''Phys. Rev. Lett. {\bf 88}, 017903 (2001).

\bibitem{Kaltenbaek2009} R.~Kaltenbaek, R.~Prevedel, M.~Aspelmeyer, and A.~Zeilinger, ``High-fidelity entanglement swapping with fully independent sources,'' Phys. Rev. A {\bf 79}, 040302 (2009).

\bibitem{Jin2015} R.-B.~Jin, M.~Takeoka, U.~Takagi, R.~Shimizu, and M.~Sasaki, ``Highly efficient entanglement swapping and teleportation at telecom wavelength,'' Sci. Rep. {\bf 5}, 9333 (2015).

\bibitem{Zhang2016} Y.~Zhang, M.~Agnew, T.~Roger, F.~S.~Roux, T.~Konrad, D.~Faccio, J.~Leach, and A.~Forbes, ``Simultaneous entanglement swapping of multiple orbital angular momentum states of light,'' Nat. Commun. {\bf 8}, 632 (2017).

\bibitem{Takei2005}
N.~Takei, H.~Yonezawa, T.~Aoki, and A.~Furusawa, ``High-Fidelity Teleportation beyond the No-Cloning Limit and Entanglement Swapping,'' Phys. Rev. Lett. {\bf 94}, 220502 (2005).

\bibitem{Riedmatten2005} H.~de~Riedmatten, I.~Marcikic, J.~A.~W.~van~Houwelingen, W.~Tittel, H.~Zbinden, and N.~Gisin, ``Long-distance entanglement swapping with photons from separated sources,'' Phys. Rev. A  {\bf 71}, 050302 (2005).

\bibitem{Takesue2009} H.~Takesue and B.~Miquel, ``Entanglement swapping using telecom-band photons generated in fibers,'' Opt. Express {\bf 17}, 10748 (2009).

\bibitem{Sangouard2011}
N.~Sangouard, B.~Sanguinetti, N.~Curtz, N.~Gisin, R.~Thew, and H.~Zbinden, ``Faithful Entanglement Swapping Based on Sum-Frequency Generation,'' Phys.~Rev.~Lett.~{\bf106}, 120403 (2011).

\bibitem{Brecht2015} B.~Brecht, D.~V.~Reddy, C.~Silberhorn, and M.~G.~Raymer, ``Photon Temporal Modes: A Complete Framework for Quantum Information Science,'' Phys. Rev. X {\bf 5}, 041017 (2015).

\bibitem{Nunn2013}
J.~Nunn, L.~J.~Wright, C.~S\"oller, L.~Zhang, I.~A.~Walmsley, and B.~J.~Smith, ``Large-alphabet time-frequency entangled quantum key distribution by means of time-to-frequency conversion,'' Opt.~Expr.~{\bf21}, 15959 (2013).

\bibitem{Donohue2014} J.~M.~Donohue, J.~Lavoie, and K.~J.~Resch, ``Ultrafast time-division demultiplexing of polarization-entangled photons,'' \prl {\bf 113}, 163602 (2014).

\bibitem{Kues2017}
M.~Kues {\em et al.}, ``On-chip generation of high-dimensional entangled quantum states and their coherent control,'' Nature {\bf 546}, 622 (2017).

\bibitem{Lukens2017} J.~M.~Lukens and P.~Lougovski, ``Frequency-encoded photonic qubits for scalable quantum information processing,'' Optica {\bf 4}, 8 (2017).

\bibitem{Karpinski2017} M.~Karpi\'nski, M.~Jachura, L.~J.~Wright, and B.~J.~Smith, ``Bandwidth manipulation of quantum light by an electro-optic time lens,'' Nat. Photon. {\bf 11}, 53 (2017).

\bibitem{Humphreys2013} P.~C.~Humphreys, B.~J.~Metcalf, J.~B.~Spring, M.~Moore, X.-M.~Jin, M.~Barbieri, W.~S.~Kolthammer, and I.~A.~Walmsley, ``Linear Optical Quantum Computing in a Single Spatial Mode'', Phys. Rev. Lett. {\bf 111}, 150501 (2013).

\bibitem{Roslund2014} J.~Roslund, R.~M.~De Ara\'ujo, S.~Jiang, C.~Fabre, and N.~Treps, ``Wavelength-multiplexed quantum networks with ultrafast frequency combs,'' Nat. Photon. {\bf 8}, 109 (2014).

\bibitem{Kowligy2014} A.~S.~Kowligy {\em et al.}, ``Quantum optical arbitrary waveform manipulation and measurement in real time,'' Opt. Express {\bf 22}, 27942 (2014).

\bibitem{Villegas2017} J.~A.~Jaramillo-Villegas, P.~Imany, O.~D.~Odele, D.~E.~Leaird, Z.-Y.~Ou, M.~Qi, and A.~M.~Weiner, ``Persistent energy-time entanglement covering multiple resonances of an on-chip biphoton frequency comb,'' Optica {\bf 4}, 655 (2017).

\bibitem{Wright2017} L.~J.~Wright, M.~Karpi\'nski, C.~S\"oller, and B.~J.~Smith, ``Spectral Shearing of Quantum Light Pulses by Electro-Optic Phase Modulation,'' Phys.~Rev.~Lett.\ {\bf 118}, 023601 (2017).

\bibitem{FiberEntanglement} Q.~Zhang, H.~Takesue, S.~W.~Nam, C.~Langrock, X.~Xie, B.~Baek, M.~M.~Fejer, and Y.~Yamamoto, ``Distribution of Time-Energy Entanglement over 100 km fiber using superconducting single-photon detectors,'' Opt. Express {\textbf 16}, 5776 (2008); T.~Inagaki, N.~Matsuda, O.~Tadanaga, M.~Asobe, and H.~Takesue, ``Entanglement distribution over 300 km of fiber,'' Opt. Express {\textbf 21}, 23241 (2013).

\bibitem{Molotkov1999}
S.~N.~Molotkov and S.~S.~Nazin, ``Photon frequency entanglement swapping,'' Phys.~Lett.~A {\bf252}, 1 (1999).

\bibitem{Guerreiro2014}
T.~Guerreiro, A.~Martin, B.~Sanguinetti, J.~S.~Pelc, C.~Langrock, M.~M.~Fejer, N.~Gisin, H.~Zbinden, N.~Sangouard, and R.~T.~Thew, ``Nonlinear Interaction between Single Photons,'' Phys. Rev. Lett. {\bf 113}, 173601 (2014).

\bibitem{Fiorentino2007} M.~Fiorentino, S.~M.~Spillane, R.~G.~Beausoleil, T.~D.~Roberts, P.~Battle, and M.~W.~Munro, ``Spontaneous parametric down-conversion in periodically poled KTP waveguides and bulk crystals,'' Opt. Express {\bf 15}, 7479 (2007).

\bibitem{Milonni1995} P.~W.~Milonni, ``Field Quantization and Radiative Processes in Dispersive Dielectric Media,'' J. Mod. Opt. {\bf 42}, 1991 (1995).

\bibitem{Blow1990a} K.~J.~Blow, R.~Loudon, S.~J.~D.~Phoenix, and T.~J.~Shepherd, ``Continuum fields in quantum optics,'' Phys. Rev. A, {\bf 42}, 4102 (1990).

\bibitem{Bierlein} Phase matching along the zyy axes corresponds to the $d_{24}$ nonlinear coefficient. Crystal structure shown in J.~D.~Bierlein and H.~Vanherzeele, ``Potassium titanyl phosphate: properties and new applications,'' J. Opt. Soc. Am. B {\bf 6}, 622 (1989); but the $d_{24}$ value we use is from H.~Vanherzeele and J.~D.~Bierlein, ``Magnitude of the nonlinear-optical coefficients of KTiOPO(4),'' Opt. Lett. {\bf 17}, 982 (1992).

\bibitem{URen2005} A.~B.~U'Ren, C.~Silberhorn, K.~Banaszek, I.~A.~Walmsley, R.~Erdmann, W.~P.~Grice, and M.~G.~Raymer, ``Generation of Pure-State Single-Photon Wavepackets by Conditional Preparation Based on Spontaneous Parametric Downconversion,'' Laser Phys.\ {\bf 15}, 146 (2005).

\bibitem{Smith2009} B.~Smith, P.~Mahou, O.~Cohen, J.~Lundeen, and I.~A.~Walmsley, ``Photon pair generation in birefringent optical fibers,'' Opt. Express {\bf 17}, 23589 (2009).

\bibitem{Davis2016}
A.~O.~C.~Davis, P.~M.~Saulnier, M.~Karpi\'nski, and B.~J.~Smith, ``Pulsed single-photon spectrograph by frequency-to-time mapping using chirped fiber Bragg gratings,'' Opt. Express {\textbf 25}, 12804 (2016).

\bibitem{Kuo2016} Note that our SFG frequencies are outside the telcom frequency range, where higher resolution has been achieved; cf. P.~S.~Kuo, T.~Gerrits, V.~B.~Verma, and S.~W.~Nam, ``Spectral correlation and interference in non-degenerate photon pairs at telecom wavelengths,'' Opt. Lett. 41, 5074 (2016).

\bibitem{Kolenderski2009}
P.~Kolenderski, W.~Wasilewski, and K.~Banaszek, ``Modeling and optimization of photon pair sources based on spontaneous parametric down-conversion,'' Phys.~Rev.~A {\bf80}, 013811 (2009).

\bibitem{Kolenderski2009b}
P.~Kolenderski and W.~Wasilewski, ``Derivation of the density matrix of a single photon produced in parametric down-conversion,'' Phys.~Rev.~A {\bf80}, 015801 (2009).

\bibitem{Vidal2002}
G.~Vidal and R.~F.~Werner, ``Computable measure of entanglement,'' Phys.~Rev.~A {\bf65}, 032314 (2002).

\bibitem{Bouwmeester1997} D.~Bouwmeester, J.-W.~Pan, K.~Mattle, M.~Eibl, H.~Weinfurter, and A.~Zeilinger, ``Experimental quantum teleportation,'' Nature {\bf 390}, 575 (1997).

\bibitem{Kato2002} K.~Kato and E.~Takaoka, ``Sellmeier and thermo-optic dispersion formulas for KTP,'' Appl. Opt. \textbf{41}, 5040 (2002).

\bibitem{Fejer1992} M.~Fejer, G.~Magel, D.~H.~Jundt, and R.~L.~Byer, ``Quasi-phase-matched second harmonic generation: tuning and tolerances,'' IEEE J.~Quant.~Electron.~{\bf 28}, 2631 (1992).

\bibitem{Taccor} The Laser Quantum Taccor is an exemplar pump laser system: \url{http://www.laserquantum.com/}

\bibitem{Chou2005}
C.~W.~Chou, H.~de~Riedmatten, D.~Felinto, S.~V.~Polyakov, S.~J.~van~Enk, and H.~J.~Kimble, ``Measurement-induced entanglement for excitation stored in remote atomic ensembles,'' Nature {\bf 438}, 828 (2005). Discussion of the coherences is in Sec. II.A of the supplementary material.

\bibitem{Lloyd08} S.~Lloyd, ``Enhanced Sensitivity of Photodetection via Quantum Illumination,'' Science {\bf321}, 1463 (2008).

\bibitem{zhuang17a} Q.~Zhuang, Z.~Zhang, and J.~H.~Shapiro, ``Entanglement-enhanced lidars for simultaneous range and velocity measurements,'' Phys. Rev. A {\bf96}, 040304(R) (2017).

\bibitem{zhuang17b} Q.~Zhuang, Z.~Zhang, and J.~H.~Shapiro, ``Optimum Mixed-State Discrimination for Noisy Entanglement-Enhanced Sensing,'' Phys. Rev. Lett. {\bf118}, 040801 (2017).

\bibitem{Reddy2015} D.~V.~Reddy, M.~G.~Raymer, and C.~J.~McKinstrie, ``Sorting photon wave packets using temporal-mode interferometry based on multiple-stage quantum frequency conversion,'' Phys. Rev. A {\textbf 91}, 012323 (2015).

\bibitem{Mathematica}
Wolfram Research, Inc., Mathematica, Version 11.1.1.0, Champaign, IL, USA (2017).

\bibitem{Matlab}
The MathWorks, Inc., MATLAB, Release 2017a, Natick, MA, USA (2017).

\bibitem{QETLAB}
Nathaniel Johnston. QETLAB: A MATLAB toolbox for quantum entanglement, version 0.9. http://www.qetlab.com, January 12, 2016. doi:10.5281/zenodo.44637

\bibitem{Rangamani2014} M.~Rangamani and M.~Rota, ``Comments on entanglement negativity in holographic field theories,'' J. High Energy Phys. {\bf2014}, 60 (2014).

\end{thebibliography}
\end{document}